\documentclass[aps,prl,twocolumn,superscriptaddress]{revtex4-1}

\usepackage{amsmath,amssymb}
\usepackage{mathrsfs}
\usepackage{graphicx}
\usepackage{color}
\usepackage[hidelinks]{hyperref}
\usepackage[normalem]{ulem}

\hypersetup{colorlinks=true, linkcolor=black, citecolor=black, urlcolor=blue}

\begin{document}

\author{Karol Gietka}
\email[Corresponding author: ]{karol.gietka@fuw.edu.pl}
\affiliation{Faculty of Physics, University of Warsaw, ul.\ Pasteura 5, 02--093 Warsaw, Poland} 
\author{Farokh Mivehvar}
\affiliation{Institut f{\"u}r Theoretische Physik, Universit{\"a}t Innsbruck, A-6020 Innsbruck, Austria}
\author{Helmut Ritsch}
\email[Corresponding author: ]{helmut.ritsch@uibk.ac.at}
\affiliation{Institut f{\"u}r Theoretische Physik, Universit{\"a}t Innsbruck, A-6020 Innsbruck, Austria}

\title{A supersolid-based gravimeter in a ring cavity}

\begin{abstract}
We propose a novel type of composite light-matter interferometer based on a supersolid-like phase of a driven Bose-Einstein condensate coupled to a pair of degenerate counterpropagating {electromagnetic} modes of an optical ring cavity. The supersolid-like condensate under the influence of the gravity drags the cavity optical potential with itself, thereby changing the relative phase of the two {cavity electromagnetic fields}. Monitoring the phase evolution of the cavity output fields thus allows for a nondestructive measurement of the gravitational acceleration. We show that the sensitivity of the proposed gravimeter exhibits Heisenberg-like scaling with respect to the atom number. As the relative phase of the cavity {fields} is insensitive to photon losses, the gravimeter is robust against these deleterious effects. For state-of-the-art experimental parameters, the relative sensitivity $\Delta g/g$ of such a gravimeter could be of the order of $10^{-10}$--$10^{-8}$ for a condensate of a half a million atoms {and interrogation time of the order of a few seconds}.
\end{abstract}

\maketitle


{\it Introduction.}---Precision measurement plays a vital role in fundamental sciences as well as technological applications. Notably, at the beginning of the twentieth century discrepancies between precise measurements and theory led to the birth of quantum mechanics~\cite{Planck1978}. Interestingly, quantum mechanics itself in turn opened an entirely new avenue in precision measurement. One of its most flourishing branches is quantum metrology, which exploits the quantum-mechanical framework to perform even more precise measurements than it is allowed by classical approaches~\cite{giovannetti2006quantum,giovannetti2011advances}. Remarkable examples include the development of precise ``gravimeters'' based on quantum mechanical effects.  

A gravimeter is an apparatus that measures the local gravitational acceleration. It allows to measure, e.g., magma build-up before volcanic eruptions, hidden hydrocarbon reserves, and Earth's tides~\cite{middlemiss2016measurement}. In addition, it also allows to test more fundamental aspects of physics such as local Lorentz invariance~\cite{flowers2017superconducting}, the isotropy of post-Newtonian gravity~\cite{muller2008atom}, and  quantum gravity~\cite{amelino2009constraining}. The current generation of gravimeters include: microelectromechanical gravimeters~\cite{middlemiss2016measurement}, free-fall gravimeters~\cite{peters1999measurement,merlet2010comparison,peters2001high,Rothleitner2009,Arnautov1983},  spring-based gravimeters~\cite{jiang20128th,lederer2009accuracy}, superconducting gravimeters~\cite{goodkind1999superconducting}, optomechanical gravimeters~\cite{armata2017quantum,qvarfort2018gravimetry}, and atom interferometers~\cite{de2008precision, cronin2009optics,PhysRevA.98.023629,abend2016atom,hu2013demonstration,Li2014}. 
 
In the above list, the atom interferometry deserves a special position because of the possibility of harnessing quantum features of many-body systems~\cite{RevModPhys.80.517}. In principle, by using entangled resources, it is possible to increase the precision of measurement over the shot-noise limit~\cite{pezze2009entanglement}. However, noise and decoherence limit the creation and use of quantum correlations~\cite{escher2011general,demkowicz2012elusive}, especially for large samples. Hence, sub-shot-noise interferometry is currently restricted to proof-of-principle experiments with atoms~\cite{Leroux2010,Appel2009,gross2012nonlinear,ockeloen2013quantum,mussel2014scalable,strobel2014fisher,lucke2011twin} as well as photons~\cite{nagata2007beating,xiang2011entanglement,kacprowicz2010experimental,afek2010high,jachura2016mode}.

In this Letter, we propose a novel type of gravimeter based on a supersolid-like state of a Bose-Einstein condensate (BEC) in an optical ring resonator~\cite{PhysRevLett.120.123601}. Supersolid is an exotic state of matter which simultaneously exhibits seemingly irreconcilable superfluid (i.e., long-range off-diagonal) and crystalline (i.e., diagonal) orders~\cite{RevModPhys.84.759}.  In other words, a supersolid is a state of matter with spontaneously broken continuous spatial transitional symmetry and internal gauge invariance. Na{\"i}vely speaking, it can be envisaged as a solid capable of a dissipationless flow. This elusive state of matter has been very recently realized in a spin-orbit-coupled BEC~\cite{li2017stripe} and a self-organized BEC inside two crossed linear cavities~\cite{leonard2017supersolid}.

\begin{figure}[b!]
    \centering

\includegraphics[width=0.43\textwidth]{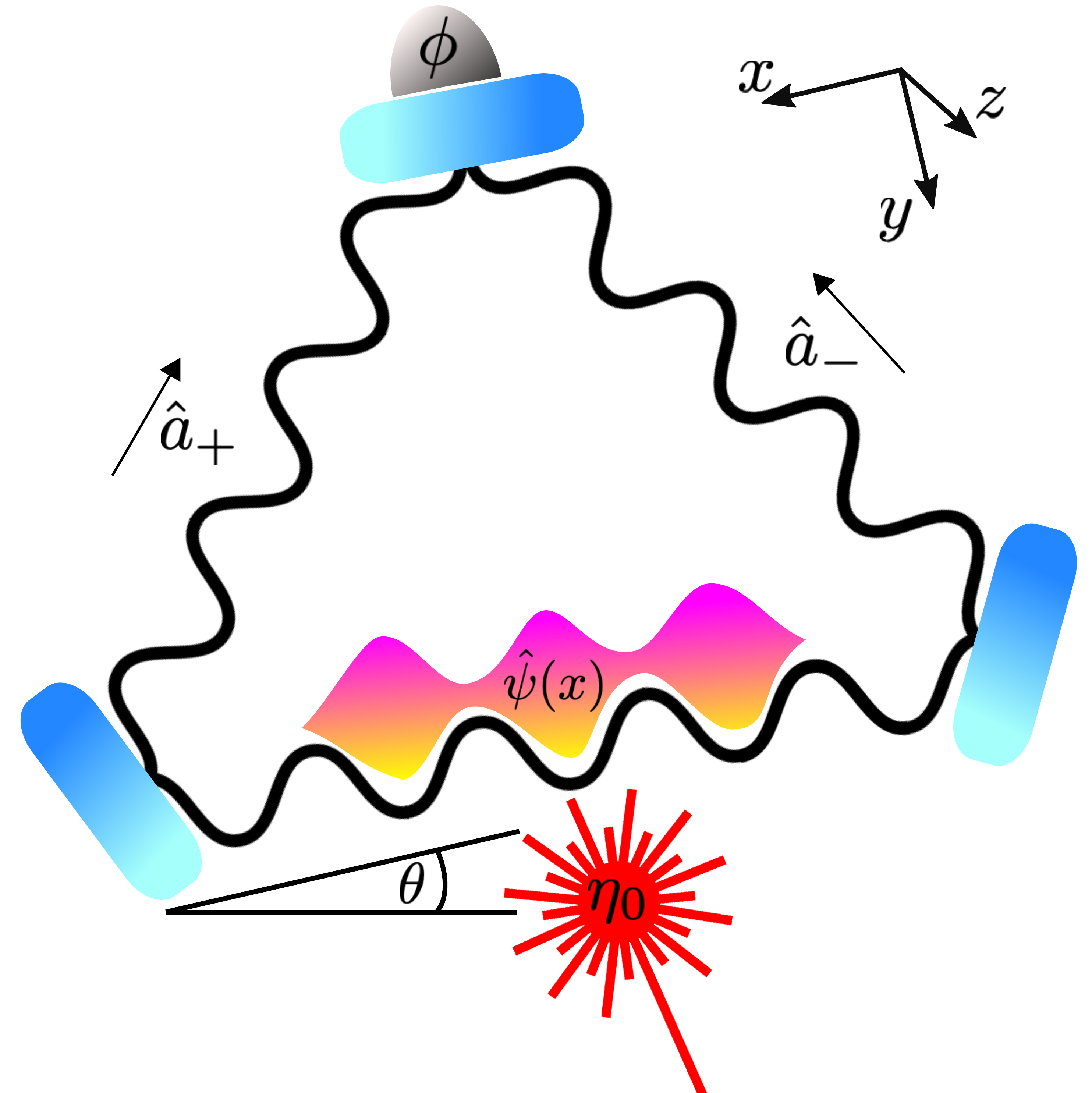}
\caption[scheme]{A schematic representation of the system. The cavity is tilted with respect to the horizontal direction by an angle~$\theta$. The relative phase $\phi$ of the two counterpropagating cavity modes $\hat{a}_{\pm}$ is measured in the cavity output to monitor the motion of the BEC along the cavity axis, described by $\hat \psi(x,t)$. The transverse pump laser is indicated by $\eta_0$.}
\label{fig:scheme}
\end{figure}

Our scheme consists of a laser-driven BEC coupled to two dynamic counterpropagating {electromagnetic} modes of a ring cavity~\cite{PhysRevLett.120.123601}, where the BEC also feels the gravitational (or any other type of) linear potential as depicted in Fig.~\ref{fig:scheme}. In the absence of the gravitational potential and above a critical pump strength, the steady state of the system is a superradiant supersolid state. More precisely, the quantum state of the system comprises a superposition of electromagnetic fields with equal amplitudes and various phases correlated with corresponding atomic density patterns. Due to quantum jumps induced by cavity photon losses~\cite{maschler2007entanglement,kramer2014self,vukics2007microscopic}, this highly entangled atom-field state collapses subsequently into a state with a certain random relative field phase and the corresponding atomic density pattern, spontaneously breaking the continuous $U(1)$ symmetry of the system and resulting in a supersolid state~\cite{SM-SS-gravimeter-2018}. The corresponding gapless Goldstone mode is the frictionless center-of-mass motion of the BEC, which drags the cavity optical lattice with itself and hence changes the relative phase of the two {electromagnetic fields of} the cavity modes {[for brevity, hereinafter referred to as just cavity (field) modes]}. Therefore, this composite atom-field dynamic can be monitored non-destructively in real time via the relative phase of the two field modes in the cavity output~\cite{Kruse2003atomic-rcoil-laser, kruse2003cold, nagorny2003optical}. That said, the deleterious aspect of photon losses does not impair the supersolid phase as photon dissipations do not affect the relative phase of the two cavity modes and the continuous $U(1)$ symmetry of the system~\cite{PhysRevLett.120.123601}. 

\begin{figure}[t!]
    \centering
\includegraphics[width=0.48\textwidth]{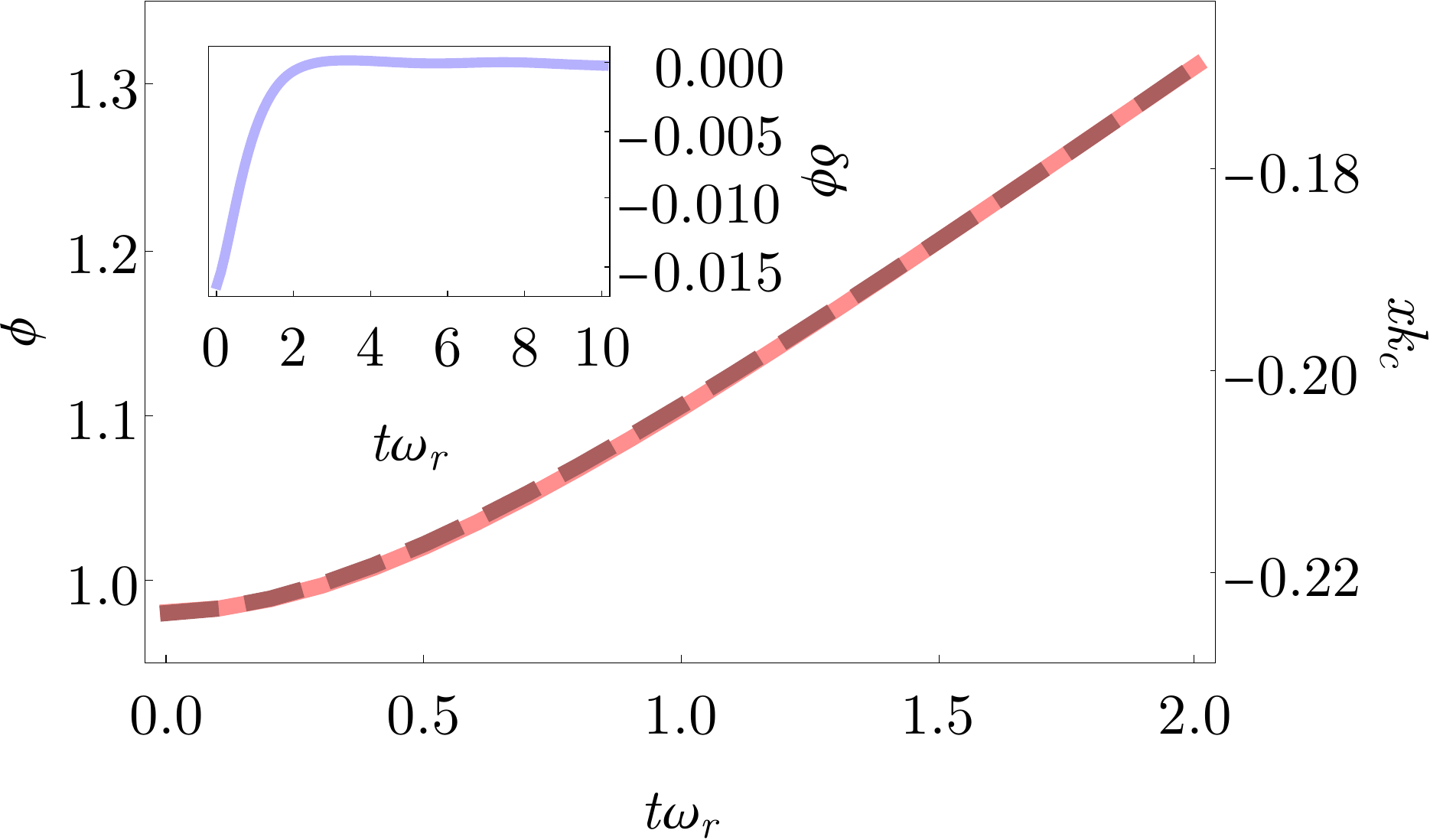}

\caption[relativephase]{The time evolution of the relative phase of the two cavity modes and the center-of-mass motion of the BEC. The solid red line corresponds to the relative phase (left axis) and the dashed red line corresponds to the center-of-mass position of the condensate (right axis), obtained numerically through the mean-field calculation. The inset shows the difference $\delta\phi$  between the numerical relative phase and the simple heuristic model of Eq.~(\ref{eq:heuristicsol}) {with fitted $\xi=0.167 \omega_r$} and $\zeta = 0.007$. The parameters are set to $(\Delta_c,\kappa)=(-8,1)\omega_r$ and $(\sqrt{N}\eta_0,NU_0,Mg\sin\theta\lambda_c)=(20,-1,10)\hbar\omega_r$, where $\omega_r \equiv \hbar k_c^2 /2M$ is the recoil frequency.}
\label{fig:relativephase}
\end{figure}

\begin{figure}[t!]
    \centering
\includegraphics[width=0.48\textwidth]{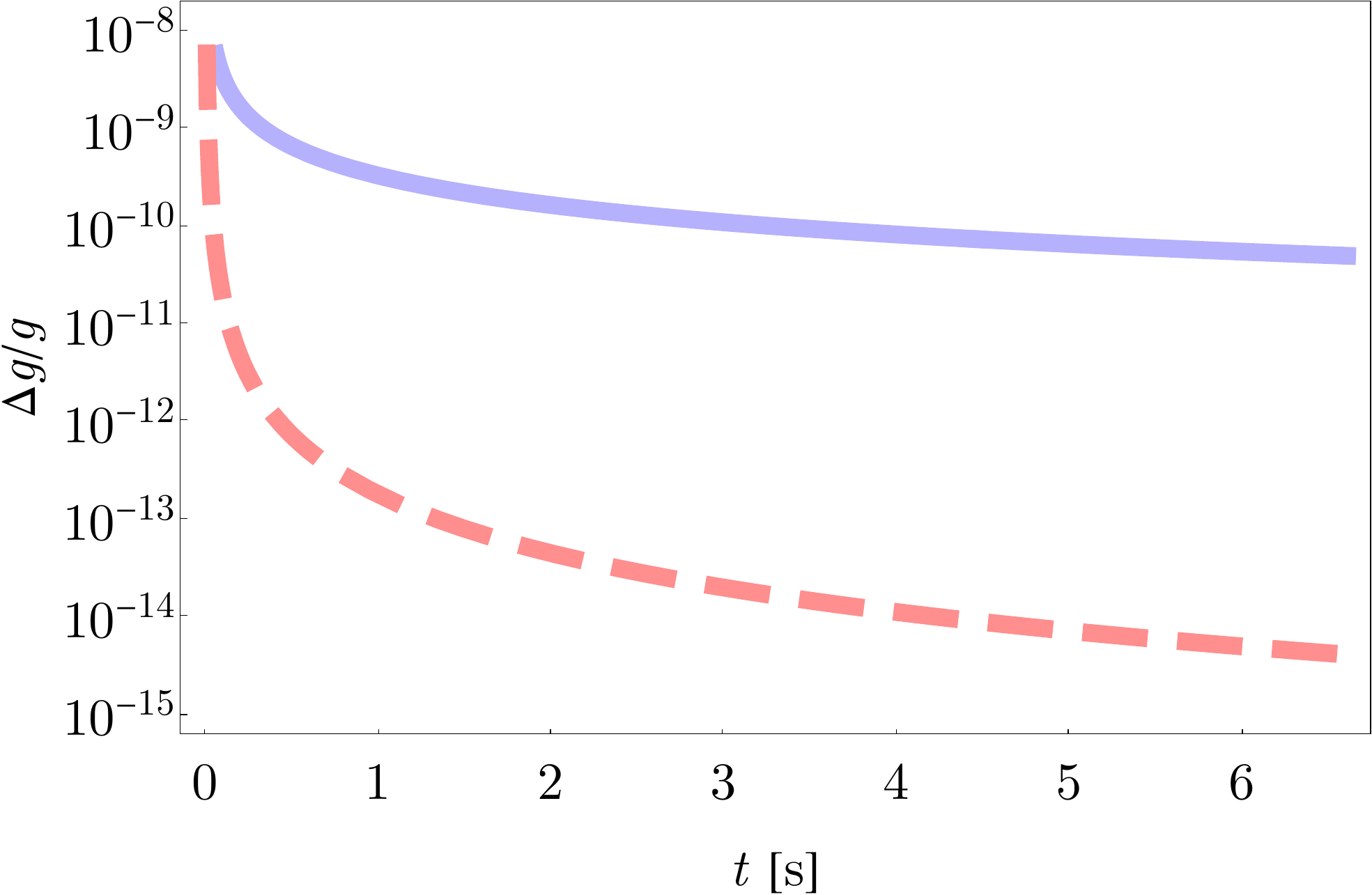}
\caption[sensitivity]{Relative sensitivity for state-of-the-art experimental parameters as a function of time. The red dashed curve presents the sensitivity for a $\kappa=0$ case (solely for the sake of comparison), while the blue solid curve takes into account the effective friction arising from the photon losses, {$\kappa=\omega_r$}. The parameters of the simulation are set to $N=5\times 10^5$, $n=2.5\times 10^{11}$, $g=9.81~{\mathrm{m}}\,{\mathrm{s}^{-2}}$, $k_c= {2 \pi}/{780}~\mathrm{nm}^{-1}$, $\xi=0.167 \omega_r$, and $\zeta = 0.007$.}
\label{fig:sensitivity}
\end{figure}

The gravitational potential breaks explicitly the continuous $U(1)$ symmetry of the system and destroys the supersolidity. As a consequence, the BEC also experiences a friction force due to photon losses. Nonetheless, the BEC still perfectly drags the cavity optical potential with itself as it moves due to the gravitational and friction forces, exhibiting a ``supersolid-like" behavior---a~direct consequence of ring geometry of the cavity. Therefore, the gravitational acceleration can be measured nondestructively in real time via the relative phase of the cavity output fields as shown in Fig.~\ref{fig:relativephase}. Due to the superradiant nature of the intra-cavity fields, their intensities are proportional to $N^2$, where $N$ is the number of atoms in the BEC. Therefore, one achieves Heisenberg-like scaling of the sensitivity by measuring the cavity fields, destructively or nondestructively, as we will demonstrate. Considering state-of-the-art experimental parameters~\cite{Kruse2003atomic-rcoil-laser, kruse2003cold,nagorny2003optical,Slama2007Rayleigh-scattering,slama2007cavity, bux2013control,schmidt2014dynamical,Culver2016,Naik2018,Cox2018ringcavity,Wolf2018,schuster2018pinning}, we obtain a relative sensitivity for measuring the gravitational acceleration $\Delta g/g$ in the order of $10^{-10}$--$10^{-8}$ {for a cycle time of the order of a few seconds}; see Fig.~\ref{fig:sensitivity}.


{\it Model.}---Consider ultracold two-level bosonic atoms which are trapped in a quasi-one dimension along one arm of a ring resonator by a tight confining potential.
This arm of the cavity makes angle $\theta$ with the horizontal direction as depicted in Fig.~\ref{fig:scheme}, so that the atoms experience the linear gravitational potential $V_g(x)=Mgx\sin\theta$ along the cavity axis. Here $M$ is the atomic mass, and $g$ is the gravitational acceleration. The atoms are transversely driven by an off-resonant pump laser which induces the transition between two relevant internal atomic states with the Rabi frequency $\Omega_0$. The atomic transition is also off-resonantly coupled to a pair of (initially empty) degenerate, counterpropagating cavity modes $\hat a_{\pm}e^{\pm ik_cx}$ with the coupling strength $\mathscr{G}_0$, where $k_c=2\pi/\lambda_c=\omega_c/c$ is the wave number of the cavity mode.

After adiabatic elimination of the atomic excited state and in the rotating frame of the pump laser, the system is described by the many-body Hamiltonian 
\begin{align} \label{eq:eff-H}
\hat{H}_{\rm eff}&=
\int \hat\psi^\dag(x)\left[-\frac{\hbar^2}{2M}\frac{\partial^2}{\partial x^2}+\hat{V}_{\rm SR}(x)+V_g(x)\right]\hat\psi(x)dx\nonumber\\
&-\hbar\Delta_c(\hat{a}_+^\dagger\hat{a}_++\hat{a}_-^\dagger\hat{a}_-),
\end{align}
where
\begin{align} \label{eq:SR-V}
 \hat{V}_{\rm SR}(x)&=
 U_0 \left[\hat{a}_+^\dag\hat{a}_++\hat{a}_-^\dag\hat{a}_-
+(\hat{a}_+^\dag\hat{a}_- e^{-2ik_cx}+\text{H.c.})\right]\nonumber\\
&+\eta_0 \left(\hat{a}_+ e^{ik_cx}+\hat{a}_- e^{-ik_cx}
+\text{H.c.}\right)
\end{align}
is the superradiant optical lattice resulting from the interference among the pump laser and the cavity modes. Here $\hat{\psi}(x)$ is the atomic annihilation field operator, and we have introduced $\Delta_c\equiv\omega_p-\omega_c$, $U_0\equiv\hbar\mathscr{G}_0^2/\Delta_a$, $\eta_0\equiv\hbar\mathscr{G}_0\Omega_0/\Delta_a$, where $\Delta_a\equiv\omega_p-\omega_a$ is the detuning between the pump-laser frequency $\omega_p$ and the atomic transition frequency $\omega_a$. The dynamic of the system is described by the Heisenberg equations of motion for the atomic field operator $i\hbar\partial_t\hat{\psi}=[\hat{\psi},\hat{H}_{\rm eff}]$ and the photonic field operators $i\hbar\partial_t\hat{a}_\pm=[\hat{a}_\pm,\hat{H}_{\rm eff}]-i\hbar\kappa\hat{a}_\pm$, where $\kappa$ is the cavity-photon loss rate. In the thermodynamic limit, quantum fluctuations are small and one can replace the field operators with the corresponding mean-field averages~\cite{piazza2013bose}: $\langle \hat\psi(x,t)\rangle=\psi(x,t)$ and $\langle\hat{a}_\pm(t)\rangle=\alpha_\pm(t)=|\alpha_\pm(t)|e^{i\phi_\pm(t)}$. We numerically solve the corresponding coupled mean-field equations using a self-consistent method~\cite{PhysRevLett.120.123601}.


\emph{Supersolid.}---When the cavity is perfectly aligned along the horizontal direction, $\theta=0$, then the gravitational potential $V_g(x)$ vanishes and the system possesses a continuous $U(1)$ symmetry, corresponding to simultaneous spatial translation $x\to x+X$ and cavity-phase rotations $\hat{a}_\pm\to\hat{a}_\pm e^{\mp i k_c X}$. Above the critical pump strength $\eta_c$ the system enters the superradiant phase, where the cavity modes are occupied and the BEC density is modulated. Although the amplitude of the two cavity modes in the superradiant state are equal, $|\alpha_+|=|\alpha_-|=|\alpha|$, their relative phase $\phi\equiv(\phi_+-\phi_-)/2$ is fixed in an arbitrary value between $0$ and $2\pi$, spontaneously breaking the $U(1)$ symmetry of the system. Corresponding to the emergent superradiant optical lattice 
$\langle \hat{V}_{\rm SR}(x) \rangle=V_{\rm SR}(x)=4U_0|\alpha|^2 \cos^2(k_cx+\phi)+4\eta_0|\alpha|\cos(k_cx+\phi)\cos(\Phi)$, 
with $\Phi\equiv(\phi_++\phi_-)/2$ being the total phase, the BEC density is modulated and forms a supersolid state. 

The gapless Goldstone mode corresponding to the spontaneously broken $U(1)$ symmetry is the center-of-mass motion of the BEC accompanied with a continuous change of the relative phase $\phi$ and hence dragging of the optical potential $V_{\rm SR}(x)$. As shown in Ref.~\cite{PhysRevLett.120.123601}, the Goldstone mode is robust against photon losses and does not acquire an imaginary part, indicating that the Goldstone mode is not damped. In other words, the density modulated BEC can move without any friction along the cavity axis dragging the superradiant optical lattice $V_{\rm SR}(x)$ with itself, explicitly exhibiting the supersolidity of the system even in the presence of photon losses.


{\it Gravimetry.}---Let us now consider the general case of $\theta\neq0$, where the atoms also feel the gravitational potential $V_g(x)$. This potential explicitly breaks the $U(1)$ symmetry of the system and destroys the supersolidity. As a consequence, a friction force is introduced into the system due to photon losses. Nevertheless, as the BEC accelerates and reaches its terminal velocity due to the interplay between the gravitational and friction forces, it still perfectly drags the cavity optical potential $V_{\rm SR}(x)$~\footnote{The modification of the cavity-field populations due to the presence of the gravitational potential are negligible for deep cavity potentials. That is, the cavity-field populations are almost the same as steady-state mean photon numbers in the absence of the gravitational potential} with itself displaying a supersolid-like behavior. This is a direct consequence of the ring geometry of the cavity, where the mirrors do not impose a static condition on locations of nodes of radiation fields, in contrary to a linear cavity. Therefore, the gravitational acceleration can be nondestructively measured via the relative phase of the cavity output fields.

Fig.~\ref{fig:relativephase} depicts the time evolution of the BEC under the free fall, $\theta=\pi/2$, obtained numerically from solving the coupled mean-field equations for $\psi$ and $\alpha_\pm$: the dashed red line (right axis) shows the position of the center-of-mass of the BEC and the solid red line (left axis) indicates the time evolution of the relative phase $\phi$. The BEC initially accelerates downward due to the gravity, dragging $V_{\rm SR}(x)$ with itself and causing the relative phase to change quadratically in time. As the time passes, the condensate, however, experiences stronger friction force and hence reaches its terminal velocity at times $t\omega_r\gtrsim 1$. Subsequently, the relative phase also evolves linearly at times $t\omega_r\gtrsim 1$, indicating that even in the presence of the gravitational potential [i.e., the absence of a perfect $U(1)$ symmetry] and dissipations the cavity optical lattice $V_{\rm SR}(x)$ follows precisely the motion of the BEC.


We now attempt to obtain a simple heuristic model to describe the time evolution of the system under the free fall and estimate the gravitational acceleration. Assuming the friction force is position independent and depends linearly on the velocity, the time evolution of the relative phase can be described by the following equation,
\begin{align} \label{eq:heuristic}
    \ddot{\phi} = \zeta g k_c - \xi \dot{\phi}
\end{align}
with the solution
\begin{align} \label{eq:heuristicsol}
    \phi(t) = \zeta \frac{ g k_c }{\xi^2} \left(e^{- \xi t} + \xi t -1  \right)+\phi_0.
\end{align}
Here $\phi_0$ is the initial relative phase, $\xi$ is the effective friction coefficient and depends on atomic and cavity parameters, most importantly on the photon decay rate $\kappa$~\cite{gangl2000cold}, and $\zeta$ is a unit-less free parameter. Both $\xi$ and $\zeta$ can be determined in an experiment via the calibration process by finding first the final velocity of the condensate $\zeta g k_c / \xi$ and then finding the characteristic time $1/\xi$. This simple model captures the time evolution of the relative phase very well as illustrated in the inset of Fig.~\ref{fig:relativephase}. Therefore, Eq.~\eqref{eq:heuristicsol} can be used to estimate the gravitational acceleration with the sensitivity given by the error propagation formula
\begin{align} \label{eq:zetun}
    \Delta g = \sqrt{\frac{1}{m} \frac{(\Delta \phi)^2}{\left(\partial_{g} \phi\right)^2}}
    = \sqrt{\frac{1}{4nm}\left[\zeta \frac{k_c}{\xi^2} \left(e^{- \xi t} + \xi t -1\right)\right]^{-2}},
\end{align}
where $m$ is the number of measurement repetitions, $n=|\alpha_+|^2+|\alpha_-|^2$ is the mean total number of photons, and we have used the phase uncertainty of coherent states to estimate the uncertainty of the relative phase $\Delta \phi \approx {1}/2\sqrt{n}$. For short times the sensitivity $\Delta g$ scales quadratically with time as opposed to typical linear scaling arising from quantum coherence, for instance, in Ramsey interferometry~\cite{PhysRevLett.111.123601}. In the context of gravimetry, a quadratic dependence of phase on time occurs for the commonly used Mach-Zehnder configuration~\cite{kasevich1991atomic,borde1989atomic} where the relative phase between the matter waves is $\phi_{\rm MZ} \propto g T^2_{\pi}$, with $2 T_{\pi}$ being the total interrogation time~\cite{kasevich1992measurement,schleich2013redshift}.

Another compelling feature of the sensitivity~(\ref{eq:zetun}) is the fact that in the superradiant regime the number of photons $n$ is proportional to $N^2$, where $N$ is the number of atoms in the BEC. Therefore, we retrieve the $SU(2)$ atomic Heisenberg scaling $1/N$ of the sensitivity~\cite{pezze2009entanglement} without explicit quantum entanglement among the atoms and the cavity fields~\cite{RevModPhys.90.035006}. Note that the quantum entanglement between the BEC and cavity fields built-up on the onset of the superradiance transition has been washed out during the quantum collapse into a product state with a well-defined relative phase and density pattern, which we have used above~\cite{SM-SS-gravimeter-2018}.

 
{\it Homodyne detection.}---To put our considerations in a more realistic context, we now focus on the homodyne detection and calculate the Fisher information. The latter sets the lower bound on the sensitivity through the Cram\'er-Rao bound $\Delta g \geq {1}/{\sqrt{mF}}$, where $F$ is the Fisher information defined as \cite{Fisher1925}
\begin{align}\label{eq:fisher}
    F = \int \mathrm{d} q \frac{1}{p(q|g)}\left[\partial_{g} p(q|g)\right]^2,
\end{align}
with $p(q|g)$ being the probability of measuring an outcome $q$ given $g$ while performing some measurement. For brevity, we restrict ourselves to homodyne measurement of combined quadrature $\hat Q=( \hat a e^{-i \gamma} + \hat a^\dagger e^{i \gamma})/2$, where $\hat a = \hat a_- +\hat a_+$ and $\gamma$ is the phase of the local oscillator. Since during the time evolution the number of photons in each mode is approximately constant and equal (confirmed by our mean-field numerical results), the homodyne signal thus reveals the relative phase: $\langle \hat Q \rangle \approx \sqrt{n} \cos\phi \cos({\Phi -\gamma})$. Now, if we choose $\gamma$ such that $\Phi -\gamma = j \pi$ with $j \in \mathbb{N}$, the probability of measuring $q$ is then given by
\begin{align}\label{eq:prob}
    p(q|g) = \sqrt{\frac{2}{\pi}}e^{-2(q-\sqrt{n} \cos\phi)^2},
\end{align}
with $\phi=\phi(t,g)$ given by Eq.~(\ref{eq:heuristicsol}). Inserting this probability into the expression for the Fisher information (\ref{eq:fisher}) yields
\begin{align}\label{eq:cfi}
    F = 4 n \left[\zeta \frac{k_c}{\xi^2} \left(e^{- \xi t} + \xi t -1\right)\sin \phi \right]^2,
\end{align}
which for optimal points in time, i.e, $\phi(t) =  (2j +1) \pi $ with $j \in \mathbb{N}$, gives the Cram{\'e}r-Rao bound on the uncertainty the same as in Eq.~(\ref{eq:zetun}). Alternatively, instead of trying to estimate the value of $g$ from a fixed point, it is more efficient to track the value of quadrature $\hat Q$ in time and fit its time dependence $\langle \hat Q(t) \rangle \approx \sqrt{n} \cos\phi(t,g)$, retrieving thus the sensitivity from Eq.~(\ref{eq:zetun}).

{\it Quantum Fisher information.}---To free ourselves from the measurement context, we calculate now the quantum Fisher information which sets the ultimate bound for the sensitivity~\cite{pezze2009entanglement}. For a pure state it can be calculated via~\cite{Braunstein1994},
\begin{align}
    F_{\rm q} = 4 (\Delta \hat h)^2,
\end{align}
where $\hat h=i[\partial_{g} \hat U(g)]\hat U^\dagger(g)$ is the generator of infinitesimal change along a trajectory parametrized by $g$.  The change of the cavity modes is effectively governed by the phase-shifting operator $\hat{U}(\phi)= \exp[- i \phi (\hat{n}_+-\hat{n}_-)]$,
with $\phi$ given by Eq.~(\ref{eq:heuristicsol}) and $\hat n_\pm=\hat{a}_\pm^\dag\hat{a}_\pm$. This yields 
$ \hat h = (k_c/\xi^2)[\exp(- \xi t) + \xi t ](\hat n_+-\hat n_-)$.

It is then straightforward to show that for coherent states quantum Fisher information is
\begin{align}
    F_{\rm q} = 4n\left[\zeta \frac{k_c}{\xi^2} \left(e^{- \xi t} + \xi t - 1\right)\right]^2,
\end{align}
which gives the Cram{\'e}r-Rao bound on the sensitivity identical to Eq.~(\ref{eq:zetun}). This should not come as a surprise since all the information about the state is stored in the relative phase~(\ref{eq:heuristicsol}). The improvement factor of 4 stems from the fact that the considered parameter-estimation scheme exploits single-mode states
\cite{pinel2013quantum}, in contrary to $SU(2)$ interferometry relying on the interference of two modes, as, for example, in Mach-Zehnder interferometry~\cite{Pezz2008}.


{\it Sensitivity.}---Finally, in order to show the potential of the supersolid-based gravimeter, we calculate the sensitivity using state-of-the-art experimental parameters. To this end, we take the parameters (number of atoms $N$ and atomic species) from recent BEC--ring-cavity experiments~\cite{Naik2018,schuster2018pinning,Wolf2018}. 
The sensitivity of the supersolid-based gravimeter containing $5\times 10^5$ $^{87}$Rb atoms is presented in Fig.~\ref{fig:sensitivity} as a solid blue curve. The sensitivity for the ideal case of frictionless motion, i.e., $\kappa=0$, is shown as a dashed red curve for the sake of comparison.

Although the results presented in Fig.~\ref{fig:sensitivity} have been obtained under the assumption that all the photons inside the cavity are measured and thus are unlikely to be achieved experimentally, they can still serve as a benchmark to give a rough estimate of a realistic sensitivity. Assuming we can extract only $10\%$ of photons from the cavity {and the measurement interval is of the order of a few seconds}, the estimated relative sensitivity might be of the order of $10^{-9}$, which is comparable with the state-of-the-art gravimeters~\cite{abend2016atom,hu2013demonstration,guzman2014high,lacoste,hardman2016simultaneous}. Moreover, taking into account the possibility of enhancing the performance of the proposed gravimeter, for example, by increasing the number of atoms and photons or by tuning other experimental parameters, the estimated per-root-Hertz relative sensitivity might be of the order of $10^{-12} \mathrm{Hz}^{-1/2}$, which is comparable with the theoretically predicted sensitivities reported in Refs.~\cite{qvarfort2018gravimetry,armata2017quantum}. However, taking into consideration systematic errors, especially arising from the fact that Eq.~(\ref{eq:heuristicsol}) is heuristic (see the inset of Fig. \ref{fig:relativephase}), and the fact that after the introduction of the gravitational potential the superradiant optical lattice is slightly modified ($\hat n_+ \approx \hat n_-$), the value of the estimated gravitational acceleration might be biased and its uncertainty may increase.


{\it Conclusions.}---We presented a novel approach for the precise measurement of the gravitational acceleration, by exploiting a supersolid-like state of a driven BEC inside an optical ring cavity. We showed the relative sensitivity of such a gravimeter is in the order of $10^{-10}$--$10^{-8}$ {for an interrogation time of the order of a few seconds}. The high sensitivity of the presented gravimeter stems from the superradiant and supersolid-like nature of the system. Since the relative phase is not affected by photon and atom losses, our proposed interferometer should be robust against these inexorable losses. In contrary to typical proposals for quantum-enhanced metrology, our interferometer does not rely on a fragile entangled state, instead exploits a robust steady state, making it to be more resilient and versatile~\cite{RevModPhys.90.035006}. Our proposed interferometer is based on state-of-the-art experiments on quantum-gas--cavity systems~\cite{Kruse2003atomic-rcoil-laser, kruse2003cold,nagorny2003optical,slama2007cavity, Slama2007Rayleigh-scattering, bux2013control,schmidt2014dynamical,Naik2018,Cox2018ringcavity,Culver2016,Wolf2018,schuster2018pinning}, and can be implemented with only minimal changes to existing experimental setups.

\begin{acknowledgments}
\emph{Acknowledgments}.---Simulations were performed using the open source QuantumOptics.jl framework in Julia~\cite{kramer2018quantumoptics}; K.G. is grateful to David Plankensteiner for related discussions. We are also grateful to F.~Piazza and Ph.~Haslinger for fruitful discussions and constructive comments. K.G.\ acknowledges financial support from the National Science Centre Poland (NCN) under the ETIUDA scholarship (2017/24/T/ST2/00161). F.M.\ is supported by the Lise-Meitner Fellowship M2438-NBL and the international FWF-ANR grant, No.\ I3964-N27. H.R.\ acknowledges support from the Austrian Science Fund FWF through No.\ I1697-N27.
\end{acknowledgments}


\begin{thebibliography}{76}%
\makeatletter
\providecommand \@ifxundefined [1]{%
 \@ifx{#1\undefined}
}%
\providecommand \@ifnum [1]{%
 \ifnum #1\expandafter \@firstoftwo
 \else \expandafter \@secondoftwo
 \fi
}%
\providecommand \@ifx [1]{%
 \ifx #1\expandafter \@firstoftwo
 \else \expandafter \@secondoftwo
 \fi
}%
\providecommand \natexlab [1]{#1}%
\providecommand \enquote  [1]{``#1''}%
\providecommand \bibnamefont  [1]{#1}%
\providecommand \bibfnamefont [1]{#1}%
\providecommand \citenamefont [1]{#1}%
\providecommand \href@noop [0]{\@secondoftwo}%
\providecommand \href [0]{\begingroup \@sanitize@url \@href}%
\providecommand \@href[1]{\@@startlink{#1}\@@href}%
\providecommand \@@href[1]{\endgroup#1\@@endlink}%
\providecommand \@sanitize@url [0]{\catcode `\\12\catcode `\$12\catcode
  `\&12\catcode `\#12\catcode `\^12\catcode `\_12\catcode `\%12\relax}%
\providecommand \@@startlink[1]{}%
\providecommand \@@endlink[0]{}%
\providecommand \url  [0]{\begingroup\@sanitize@url \@url }%
\providecommand \@url [1]{\endgroup\@href {#1}{\urlprefix }}%
\providecommand \urlprefix  [0]{URL }%
\providecommand \Eprint [0]{\href }%
\providecommand \doibase [0]{http://dx.doi.org/}%
\providecommand \selectlanguage [0]{\@gobble}%
\providecommand \bibinfo  [0]{\@secondoftwo}%
\providecommand \bibfield  [0]{\@secondoftwo}%
\providecommand \translation [1]{[#1]}%
\providecommand \BibitemOpen [0]{}%
\providecommand \bibitemStop [0]{}%
\providecommand \bibitemNoStop [0]{.\EOS\space}%
\providecommand \EOS [0]{\spacefactor3000\relax}%
\providecommand \BibitemShut  [1]{\csname bibitem#1\endcsname}%
\let\auto@bib@innerbib\@empty
\bibitem [{\citenamefont {Sch\"{o}pf}(1978)}]{Planck1978}%
  \BibitemOpen
  \bibinfo {editor} {\bibfnamefont {H.-G.}\ \bibnamefont {Sch\"{o}pf}},\ ed.,\
  \href {https://doi.org/10.1007/978-3-663-13885-3} {\emph {\bibinfo {title}
  {Von Kirchhoff bis Planck}}}\ (\bibinfo  {publisher} {Vieweg$+$Teubner
  Verlag},\ \bibinfo {year} {1978})\BibitemShut {NoStop}%
\bibitem [{\citenamefont {Giovannetti}\ \emph {et~al.}(2006)\citenamefont
  {Giovannetti}, \citenamefont {Lloyd},\ and\ \citenamefont
  {Maccone}}]{giovannetti2006quantum}%
  \BibitemOpen
  \bibfield  {author} {\bibinfo {author} {\bibfnamefont {V.}~\bibnamefont
  {Giovannetti}}, \bibinfo {author} {\bibfnamefont {S.}~\bibnamefont {Lloyd}},
  \ and\ \bibinfo {author} {\bibfnamefont {L.}~\bibnamefont {Maccone}},\ }\href
  {https://link.aps.org/doi/10.1103/PhysRevLett.96.010401} {\bibfield
  {journal} {\bibinfo  {journal} {Phys. Rev. Lett.}\ }\textbf {\bibinfo
  {volume} {96}},\ \bibinfo {pages} {010401} (\bibinfo {year}
  {2006})}\BibitemShut {NoStop}%
\bibitem [{\citenamefont {Giovannetti}\ \emph {et~al.}(2011)\citenamefont
  {Giovannetti}, \citenamefont {Lloyd},\ and\ \citenamefont
  {Maccone}}]{giovannetti2011advances}%
  \BibitemOpen
  \bibfield  {author} {\bibinfo {author} {\bibfnamefont {V.}~\bibnamefont
  {Giovannetti}}, \bibinfo {author} {\bibfnamefont {S.}~\bibnamefont {Lloyd}},
  \ and\ \bibinfo {author} {\bibfnamefont {L.}~\bibnamefont {Maccone}},\ }\href
  {https://doi.org/10.1038/nphoton.2011.35} {\bibfield  {journal} {\bibinfo
  {journal} {Nature Photonics}\ }\textbf {\bibinfo {volume} {5}},\ \bibinfo
  {pages} {222} (\bibinfo {year} {2011})}\BibitemShut {NoStop}%
\bibitem [{\citenamefont {Middlemiss}\ \emph {et~al.}(2016)\citenamefont
  {Middlemiss}, \citenamefont {Samarelli}, \citenamefont {Paul}, \citenamefont
  {Hough}, \citenamefont {Rowan},\ and\ \citenamefont
  {Hammond}}]{middlemiss2016measurement}%
  \BibitemOpen
  \bibfield  {author} {\bibinfo {author} {\bibfnamefont {R.~P.}\ \bibnamefont
  {Middlemiss}}, \bibinfo {author} {\bibfnamefont {A.}~\bibnamefont
  {Samarelli}}, \bibinfo {author} {\bibfnamefont {D.~J.}\ \bibnamefont {Paul}},
  \bibinfo {author} {\bibfnamefont {J.}~\bibnamefont {Hough}}, \bibinfo
  {author} {\bibfnamefont {S.}~\bibnamefont {Rowan}}, \ and\ \bibinfo {author}
  {\bibfnamefont {G.~D.}\ \bibnamefont {Hammond}},\ }\href
  {https://doi.org/10.1038/nature17397} {\bibfield  {journal} {\bibinfo
  {journal} {Nature}\ }\textbf {\bibinfo {volume} {531}},\ \bibinfo {pages}
  {614} (\bibinfo {year} {2016})}\BibitemShut {NoStop}%
\bibitem [{\citenamefont {Flowers}\ \emph {et~al.}(2017)\citenamefont
  {Flowers}, \citenamefont {Goodge},\ and\ \citenamefont
  {Tasson}}]{flowers2017superconducting}%
  \BibitemOpen
  \bibfield  {author} {\bibinfo {author} {\bibfnamefont {N.~A.}\ \bibnamefont
  {Flowers}}, \bibinfo {author} {\bibfnamefont {C.}~\bibnamefont {Goodge}}, \
  and\ \bibinfo {author} {\bibfnamefont {J.~D.}\ \bibnamefont {Tasson}},\
  }\href {https://link.aps.org/doi/10.1103/PhysRevLett.119.201101} {\bibfield
  {journal} {\bibinfo  {journal} {Phys. Rev. Lett.}\ }\textbf {\bibinfo
  {volume} {119}},\ \bibinfo {pages} {201101} (\bibinfo {year}
  {2017})}\BibitemShut {NoStop}%
\bibitem [{\citenamefont {M\"uller}\ \emph {et~al.}(2008)\citenamefont
  {M\"uller}, \citenamefont {Chiow}, \citenamefont {Herrmann}, \citenamefont
  {Chu},\ and\ \citenamefont {Chung}}]{muller2008atom}%
  \BibitemOpen
  \bibfield  {author} {\bibinfo {author} {\bibfnamefont {H.}~\bibnamefont
  {M\"uller}}, \bibinfo {author} {\bibfnamefont {S.-w.}\ \bibnamefont {Chiow}},
  \bibinfo {author} {\bibfnamefont {S.}~\bibnamefont {Herrmann}}, \bibinfo
  {author} {\bibfnamefont {S.}~\bibnamefont {Chu}}, \ and\ \bibinfo {author}
  {\bibfnamefont {K.-Y.}\ \bibnamefont {Chung}},\ }\href
  {https://link.aps.org/doi/10.1103/PhysRevLett.100.031101} {\bibfield
  {journal} {\bibinfo  {journal} {Phys. Rev. Lett.}\ }\textbf {\bibinfo
  {volume} {100}},\ \bibinfo {pages} {031101} (\bibinfo {year}
  {2008})}\BibitemShut {NoStop}%
\bibitem [{\citenamefont {Amelino-Camelia}\ \emph {et~al.}(2009)\citenamefont
  {Amelino-Camelia}, \citenamefont {L\"ammerzahl}, \citenamefont {Mercati},\
  and\ \citenamefont {Tino}}]{amelino2009constraining}%
  \BibitemOpen
  \bibfield  {author} {\bibinfo {author} {\bibfnamefont {G.}~\bibnamefont
  {Amelino-Camelia}}, \bibinfo {author} {\bibfnamefont {C.}~\bibnamefont
  {L\"ammerzahl}}, \bibinfo {author} {\bibfnamefont {F.}~\bibnamefont
  {Mercati}}, \ and\ \bibinfo {author} {\bibfnamefont {G.~M.}\ \bibnamefont
  {Tino}},\ }\href {https://link.aps.org/doi/10.1103/PhysRevLett.103.171302}
  {\bibfield  {journal} {\bibinfo  {journal} {Phys. Rev. Lett.}\ }\textbf
  {\bibinfo {volume} {103}},\ \bibinfo {pages} {171302} (\bibinfo {year}
  {2009})}\BibitemShut {NoStop}%
\bibitem [{\citenamefont {Peters}\ \emph {et~al.}(1999)\citenamefont {Peters},
  \citenamefont {Chung},\ and\ \citenamefont {Chu}}]{peters1999measurement}%
  \BibitemOpen
  \bibfield  {author} {\bibinfo {author} {\bibfnamefont {A.}~\bibnamefont
  {Peters}}, \bibinfo {author} {\bibfnamefont {K.~Y.}\ \bibnamefont {Chung}}, \
  and\ \bibinfo {author} {\bibfnamefont {S.}~\bibnamefont {Chu}},\ }\href
  {https://doi.org/10.1038/23655} {\bibfield  {journal} {\bibinfo  {journal}
  {Nature}\ }\textbf {\bibinfo {volume} {400}},\ \bibinfo {pages} {849}
  (\bibinfo {year} {1999})}\BibitemShut {NoStop}%
\bibitem [{\citenamefont {Merlet}\ \emph {et~al.}(2010)\citenamefont {Merlet},
  \citenamefont {Bodart}, \citenamefont {Malossi}, \citenamefont {Landragin},
  \citenamefont {Santos}, \citenamefont {Gitlein},\ and\ \citenamefont
  {Timmen}}]{merlet2010comparison}%
  \BibitemOpen
  \bibfield  {author} {\bibinfo {author} {\bibfnamefont {S.}~\bibnamefont
  {Merlet}}, \bibinfo {author} {\bibfnamefont {Q.}~\bibnamefont {Bodart}},
  \bibinfo {author} {\bibfnamefont {N.}~\bibnamefont {Malossi}}, \bibinfo
  {author} {\bibfnamefont {A.}~\bibnamefont {Landragin}}, \bibinfo {author}
  {\bibfnamefont {F.~P.~D.}\ \bibnamefont {Santos}}, \bibinfo {author}
  {\bibfnamefont {O.}~\bibnamefont {Gitlein}}, \ and\ \bibinfo {author}
  {\bibfnamefont {L.}~\bibnamefont {Timmen}},\ }\href
  {https://doi.org/10.1088/0026-1394/47/4/l01} {\bibfield  {journal} {\bibinfo
  {journal} {Metrologia}\ }\textbf {\bibinfo {volume} {47}},\ \bibinfo {pages}
  {L9} (\bibinfo {year} {2010})}\BibitemShut {NoStop}%
\bibitem [{\citenamefont {Peters}\ \emph {et~al.}(2001)\citenamefont {Peters},
  \citenamefont {Chung},\ and\ \citenamefont {Chu}}]{peters2001high}%
  \BibitemOpen
  \bibfield  {author} {\bibinfo {author} {\bibfnamefont {A.}~\bibnamefont
  {Peters}}, \bibinfo {author} {\bibfnamefont {K.~Y.}\ \bibnamefont {Chung}}, \
  and\ \bibinfo {author} {\bibfnamefont {S.}~\bibnamefont {Chu}},\ }\href
  {https://doi.org/10.1088/0026-1394/38/1/4} {\bibfield  {journal} {\bibinfo
  {journal} {Metrologia}\ }\textbf {\bibinfo {volume} {38}},\ \bibinfo {pages}
  {25} (\bibinfo {year} {2001})}\BibitemShut {NoStop}%
\bibitem [{\citenamefont {Rothleitner}\ \emph {et~al.}(2009)\citenamefont
  {Rothleitner}, \citenamefont {Svitlov}, \citenamefont
  {M{\'{e}}rim{\`{e}}che}, \citenamefont {Hu},\ and\ \citenamefont
  {Wang}}]{Rothleitner2009}%
  \BibitemOpen
  \bibfield  {author} {\bibinfo {author} {\bibfnamefont {C.}~\bibnamefont
  {Rothleitner}}, \bibinfo {author} {\bibfnamefont {S.}~\bibnamefont
  {Svitlov}}, \bibinfo {author} {\bibfnamefont {H.}~\bibnamefont
  {M{\'{e}}rim{\`{e}}che}}, \bibinfo {author} {\bibfnamefont {H.}~\bibnamefont
  {Hu}}, \ and\ \bibinfo {author} {\bibfnamefont {L.~J.}\ \bibnamefont
  {Wang}},\ }\href {\doibase 10.1088/0026-1394/46/3/017} {\bibfield  {journal}
  {\bibinfo  {journal} {Metrologia}\ }\textbf {\bibinfo {volume} {46}},\
  \bibinfo {pages} {283} (\bibinfo {year} {2009})}\BibitemShut {NoStop}%
\bibitem [{\citenamefont {Arnautov}\ \emph {et~al.}(1983)\citenamefont
  {Arnautov}, \citenamefont {Boulanger}, \citenamefont {Kalish}, \citenamefont
  {Koronkevitch}, \citenamefont {Stus},\ and\ \citenamefont
  {Tarasyuk}}]{Arnautov1983}%
  \BibitemOpen
  \bibfield  {author} {\bibinfo {author} {\bibfnamefont {G.~P.}\ \bibnamefont
  {Arnautov}}, \bibinfo {author} {\bibfnamefont {Y.~D.}\ \bibnamefont
  {Boulanger}}, \bibinfo {author} {\bibfnamefont {E.~N.}\ \bibnamefont
  {Kalish}}, \bibinfo {author} {\bibfnamefont {V.~P.}\ \bibnamefont
  {Koronkevitch}}, \bibinfo {author} {\bibfnamefont {Y.~F.}\ \bibnamefont
  {Stus}}, \ and\ \bibinfo {author} {\bibfnamefont {V.~G.}\ \bibnamefont
  {Tarasyuk}},\ }\href {\doibase 10.1088/0026-1394/19/2/001} {\bibfield
  {journal} {\bibinfo  {journal} {Metrologia}\ }\textbf {\bibinfo {volume}
  {19}},\ \bibinfo {pages} {49} (\bibinfo {year} {1983})}\BibitemShut {NoStop}%
\bibitem [{\citenamefont {Jiang}\ \emph {et~al.}(2012)\citenamefont {Jiang},
  \citenamefont {P{\'{a}}link{\'{a}}{\v{s}}}, \citenamefont {Arias},
  \citenamefont {Liard}, \citenamefont {Merlet}, \citenamefont {Wilmes},
  \citenamefont {Vitushkin}, \citenamefont {Robertsson}, \citenamefont
  {Tisserand}, \citenamefont {Santos},\ and\ \citenamefont {{\it{et
  al.}}}}]{jiang20128th}%
  \BibitemOpen
  \bibfield  {author} {\bibinfo {author} {\bibfnamefont {Z.}~\bibnamefont
  {Jiang}}, \bibinfo {author} {\bibfnamefont {V.}~\bibnamefont
  {P{\'{a}}link{\'{a}}{\v{s}}}}, \bibinfo {author} {\bibfnamefont {F.~E.}\
  \bibnamefont {Arias}}, \bibinfo {author} {\bibfnamefont {J.}~\bibnamefont
  {Liard}}, \bibinfo {author} {\bibfnamefont {S.}~\bibnamefont {Merlet}},
  \bibinfo {author} {\bibfnamefont {H.}~\bibnamefont {Wilmes}}, \bibinfo
  {author} {\bibfnamefont {L.}~\bibnamefont {Vitushkin}}, \bibinfo {author}
  {\bibfnamefont {L.}~\bibnamefont {Robertsson}}, \bibinfo {author}
  {\bibfnamefont {L.}~\bibnamefont {Tisserand}}, \bibinfo {author}
  {\bibfnamefont {F.~P.~D.}\ \bibnamefont {Santos}}, \ and\ \bibinfo {author}
  {\bibnamefont {{\it{et al.}}}},\ }\href
  {https://doi.org/10.1088/0026-1394/49/6/666} {\bibfield  {journal} {\bibinfo
  {journal} {Metrologia}\ }\textbf {\bibinfo {volume} {49}},\ \bibinfo {pages}
  {666} (\bibinfo {year} {2012})}\BibitemShut {NoStop}%
\bibitem [{\citenamefont {Lederer}(2009)}]{lederer2009accuracy}%
  \BibitemOpen
  \bibfield  {author} {\bibinfo {author} {\bibfnamefont {M.}~\bibnamefont
  {Lederer}},\ }\href
  {https://www.irsm.cas.cz/materialy/acta_content/2009_03/17_Lederer.pdf}
  {\bibfield  {journal} {\bibinfo  {journal} {Acta Geodyn. Geomater}\ }\textbf
  {\bibinfo {volume} {6}},\ \bibinfo {pages} {155} (\bibinfo {year}
  {2009})}\BibitemShut {NoStop}%
\bibitem [{\citenamefont {Goodkind}(1999)}]{goodkind1999superconducting}%
  \BibitemOpen
  \bibfield  {author} {\bibinfo {author} {\bibfnamefont {J.~M.}\ \bibnamefont
  {Goodkind}},\ }\href {https://doi.org/10.1063/1.1150092} {\bibfield
  {journal} {\bibinfo  {journal} {Review of Scientific Instruments}\ }\textbf
  {\bibinfo {volume} {70}},\ \bibinfo {pages} {4131} (\bibinfo {year}
  {1999})}\BibitemShut {NoStop}%
\bibitem [{\citenamefont {Armata}\ \emph {et~al.}(2017)\citenamefont {Armata},
  \citenamefont {Latmiral}, \citenamefont {Plato},\ and\ \citenamefont
  {Kim}}]{armata2017quantum}%
  \BibitemOpen
  \bibfield  {author} {\bibinfo {author} {\bibfnamefont {F.}~\bibnamefont
  {Armata}}, \bibinfo {author} {\bibfnamefont {L.}~\bibnamefont {Latmiral}},
  \bibinfo {author} {\bibfnamefont {A.~D.~K.}\ \bibnamefont {Plato}}, \ and\
  \bibinfo {author} {\bibfnamefont {M.~S.}\ \bibnamefont {Kim}},\ }\href
  {https://link.aps.org/doi/10.1103/PhysRevA.96.043824} {\bibfield  {journal}
  {\bibinfo  {journal} {Phys. Rev. A}\ }\textbf {\bibinfo {volume} {96}},\
  \bibinfo {pages} {043824} (\bibinfo {year} {2017})}\BibitemShut {NoStop}%
\bibitem [{\citenamefont {Qvarfort}\ \emph {et~al.}(2018)\citenamefont
  {Qvarfort}, \citenamefont {Serafini}, \citenamefont {Barker},\ and\
  \citenamefont {Bose}}]{qvarfort2018gravimetry}%
  \BibitemOpen
  \bibfield  {author} {\bibinfo {author} {\bibfnamefont {S.}~\bibnamefont
  {Qvarfort}}, \bibinfo {author} {\bibfnamefont {A.}~\bibnamefont {Serafini}},
  \bibinfo {author} {\bibfnamefont {P.~F.}\ \bibnamefont {Barker}}, \ and\
  \bibinfo {author} {\bibfnamefont {S.}~\bibnamefont {Bose}},\ }\href
  {https://doi.org/10.1038/s41467-018-06037-z} {\bibfield  {journal} {\bibinfo
  {journal} {Nature Communications}\ }\textbf {\bibinfo {volume} {9}}, {\bibinfo {pages} {3690}} (\bibinfo
  {year} {2018})}\BibitemShut {NoStop}%
\bibitem [{\citenamefont {de~Angelis}\ \emph {et~al.}(2008)\citenamefont
  {de~Angelis}, \citenamefont {Bertoldi}, \citenamefont {Cacciapuoti},
  \citenamefont {Giorgini}, \citenamefont {Lamporesi}, \citenamefont
  {Prevedelli}, \citenamefont {Saccorotti}, \citenamefont {Sorrentino},\ and\
  \citenamefont {Tino}}]{de2008precision}%
  \BibitemOpen
  \bibfield  {author} {\bibinfo {author} {\bibfnamefont {M.}~\bibnamefont
  {de~Angelis}}, \bibinfo {author} {\bibfnamefont {A.}~\bibnamefont
  {Bertoldi}}, \bibinfo {author} {\bibfnamefont {L.}~\bibnamefont
  {Cacciapuoti}}, \bibinfo {author} {\bibfnamefont {A.}~\bibnamefont
  {Giorgini}}, \bibinfo {author} {\bibfnamefont {G.}~\bibnamefont {Lamporesi}},
  \bibinfo {author} {\bibfnamefont {M.}~\bibnamefont {Prevedelli}}, \bibinfo
  {author} {\bibfnamefont {G.}~\bibnamefont {Saccorotti}}, \bibinfo {author}
  {\bibfnamefont {F.}~\bibnamefont {Sorrentino}}, \ and\ \bibinfo {author}
  {\bibfnamefont {G.~M.}\ \bibnamefont {Tino}},\ }\href
  {https://doi.org/10.1088/0957-0233/20/2/022001} {\bibfield  {journal}
  {\bibinfo  {journal} {Measurement Science and Technology}\ }\textbf {\bibinfo
  {volume} {20}},\ \bibinfo {pages} {022001} (\bibinfo {year}
  {2008})}\BibitemShut {NoStop}%
\bibitem [{\citenamefont {Cronin}\ \emph {et~al.}(2009)\citenamefont {Cronin},
  \citenamefont {Schmiedmayer},\ and\ \citenamefont
  {Pritchard}}]{cronin2009optics}%
  \BibitemOpen
  \bibfield  {author} {\bibinfo {author} {\bibfnamefont {A.~D.}\ \bibnamefont
  {Cronin}}, \bibinfo {author} {\bibfnamefont {J.}~\bibnamefont
  {Schmiedmayer}}, \ and\ \bibinfo {author} {\bibfnamefont {D.~E.}\
  \bibnamefont {Pritchard}},\ }\href
  {https://link.aps.org/doi/10.1103/RevModPhys.81.1051} {\bibfield  {journal}
  {\bibinfo  {journal} {Rev. Mod. Phys.}\ }\textbf {\bibinfo {volume} {81}},\
  \bibinfo {pages} {1051} (\bibinfo {year} {2009})}\BibitemShut {NoStop}%
\bibitem [{\citenamefont {Kritsotakis}\ \emph {et~al.}(2018)\citenamefont
  {Kritsotakis}, \citenamefont {Szigeti}, \citenamefont {Dunningham},\ and\
  \citenamefont {Haine}}]{PhysRevA.98.023629}%
  \BibitemOpen
  \bibfield  {author} {\bibinfo {author} {\bibfnamefont {M.}~\bibnamefont
  {Kritsotakis}}, \bibinfo {author} {\bibfnamefont {S.~S.}\ \bibnamefont
  {Szigeti}}, \bibinfo {author} {\bibfnamefont {J.~A.}\ \bibnamefont
  {Dunningham}}, \ and\ \bibinfo {author} {\bibfnamefont {S.~A.}\ \bibnamefont
  {Haine}},\ }\href {https://link.aps.org/doi/10.1103/PhysRevA.98.023629}
  {\bibfield  {journal} {\bibinfo  {journal} {Phys. Rev. A}\ }\textbf {\bibinfo
  {volume} {98}},\ \bibinfo {pages} {023629} (\bibinfo {year}
  {2018})}\BibitemShut {NoStop}%
\bibitem [{\citenamefont {Abend}\ \emph {et~al.}(2016)\citenamefont {Abend},
  \citenamefont {Gebbe}, \citenamefont {Gersemann}, \citenamefont {Ahlers},
  \citenamefont {M\"untinga}, \citenamefont {Giese}, \citenamefont {Gaaloul},
  \citenamefont {Schubert}, \citenamefont {L\"ammerzahl}, \citenamefont
  {Ertmer}, \citenamefont {Schleich},\ and\ \citenamefont
  {Rasel}}]{abend2016atom}%
  \BibitemOpen
  \bibfield  {author} {\bibinfo {author} {\bibfnamefont {S.}~\bibnamefont
  {Abend}}, \bibinfo {author} {\bibfnamefont {M.}~\bibnamefont {Gebbe}},
  \bibinfo {author} {\bibfnamefont {M.}~\bibnamefont {Gersemann}}, \bibinfo
  {author} {\bibfnamefont {H.}~\bibnamefont {Ahlers}}, \bibinfo {author}
  {\bibfnamefont {H.}~\bibnamefont {M\"untinga}}, \bibinfo {author}
  {\bibfnamefont {E.}~\bibnamefont {Giese}}, \bibinfo {author} {\bibfnamefont
  {N.}~\bibnamefont {Gaaloul}}, \bibinfo {author} {\bibfnamefont
  {C.}~\bibnamefont {Schubert}}, \bibinfo {author} {\bibfnamefont
  {C.}~\bibnamefont {L\"ammerzahl}}, \bibinfo {author} {\bibfnamefont
  {W.}~\bibnamefont {Ertmer}}, \bibinfo {author} {\bibfnamefont {W.~P.}\
  \bibnamefont {Schleich}}, \ and\ \bibinfo {author} {\bibfnamefont {E.~M.}\
  \bibnamefont {Rasel}},\ }\href
  {https://link.aps.org/doi/10.1103/PhysRevLett.117.203003} {\bibfield
  {journal} {\bibinfo  {journal} {Phys. Rev. Lett.}\ }\textbf {\bibinfo
  {volume} {117}},\ \bibinfo {pages} {203003} (\bibinfo {year}
  {2016})}\BibitemShut {NoStop}%
\bibitem [{\citenamefont {Hu}\ \emph {et~al.}(2013)\citenamefont {Hu},
  \citenamefont {Sun}, \citenamefont {Duan}, \citenamefont {Zhou},
  \citenamefont {Chen}, \citenamefont {Zhan}, \citenamefont {Zhang},\ and\
  \citenamefont {Luo}}]{hu2013demonstration}%
  \BibitemOpen
  \bibfield  {author} {\bibinfo {author} {\bibfnamefont {Z.-K.}\ \bibnamefont
  {Hu}}, \bibinfo {author} {\bibfnamefont {B.-L.}\ \bibnamefont {Sun}},
  \bibinfo {author} {\bibfnamefont {X.-C.}\ \bibnamefont {Duan}}, \bibinfo
  {author} {\bibfnamefont {M.-K.}\ \bibnamefont {Zhou}}, \bibinfo {author}
  {\bibfnamefont {L.-L.}\ \bibnamefont {Chen}}, \bibinfo {author}
  {\bibfnamefont {S.}~\bibnamefont {Zhan}}, \bibinfo {author} {\bibfnamefont
  {Q.-Z.}\ \bibnamefont {Zhang}}, \ and\ \bibinfo {author} {\bibfnamefont
  {J.}~\bibnamefont {Luo}},\ }\href
  {https://link.aps.org/doi/10.1103/PhysRevA.88.043610} {\bibfield  {journal}
  {\bibinfo  {journal} {Phys. Rev. A}\ }\textbf {\bibinfo {volume} {88}},\
  \bibinfo {pages} {043610} (\bibinfo {year} {2013})}\BibitemShut {NoStop}%
\bibitem [{\citenamefont {Li}\ \emph {et~al.}(2014)\citenamefont {Li},
  \citenamefont {He},\ and\ \citenamefont {Smerzi}}]{Li2014}%
  \BibitemOpen
  \bibfield  {author} {\bibinfo {author} {\bibfnamefont {W.D.}~\bibnamefont
  {Li}}, \bibinfo {author} {\bibfnamefont {T.}~\bibnamefont {He}}, \ and\
  \bibinfo {author} {\bibfnamefont {A.}~\bibnamefont {Smerzi}},\ }\href
  {\doibase 10.1103/physrevlett.113.023003} {\bibfield  {journal} {\bibinfo
  {journal} {Phys. Rev. Lett.}\ }\textbf {\bibinfo {volume} {113}}, \bibinfo {pages} {023003}
  (\bibinfo {year} {2014})}\BibitemShut
  {NoStop}%
\bibitem [{\citenamefont {Amico}\ \emph {et~al.}(2008)\citenamefont {Amico},
  \citenamefont {Fazio}, \citenamefont {Osterloh},\ and\ \citenamefont
  {Vedral}}]{RevModPhys.80.517}%
  \BibitemOpen
  \bibfield  {author} {\bibinfo {author} {\bibfnamefont {L.}~\bibnamefont
  {Amico}}, \bibinfo {author} {\bibfnamefont {R.}~\bibnamefont {Fazio}},
  \bibinfo {author} {\bibfnamefont {A.}~\bibnamefont {Osterloh}}, \ and\
  \bibinfo {author} {\bibfnamefont {V.}~\bibnamefont {Vedral}},\ }\href
  {https://link.aps.org/doi/10.1103/RevModPhys.80.517} {\bibfield  {journal}
  {\bibinfo  {journal} {Rev. Mod. Phys.}\ }\textbf {\bibinfo {volume} {80}},\
  \bibinfo {pages} {517} (\bibinfo {year} {2008})}\BibitemShut {NoStop}%
\bibitem [{\citenamefont {Pezz\'e}\ and\ \citenamefont
  {Smerzi}(2009)}]{pezze2009entanglement}%
  \BibitemOpen
  \bibfield  {author} {\bibinfo {author} {\bibfnamefont {L.}~\bibnamefont
  {Pezz\'e}}\ and\ \bibinfo {author} {\bibfnamefont {A.}~\bibnamefont
  {Smerzi}},\ }\href {https://link.aps.org/doi/10.1103/PhysRevLett.102.100401}
  {\bibfield  {journal} {\bibinfo  {journal} {Phys. Rev. Lett.}\ }\textbf
  {\bibinfo {volume} {102}},\ \bibinfo {pages} {100401} (\bibinfo {year}
  {2009})}\BibitemShut {NoStop}%
\bibitem [{\citenamefont {Escher}\ \emph {et~al.}(2011)\citenamefont {Escher},
  \citenamefont {de~Matos~Filho},\ and\ \citenamefont
  {Davidovich}}]{escher2011general}%
  \BibitemOpen
  \bibfield  {author} {\bibinfo {author} {\bibfnamefont {B.~M.}\ \bibnamefont
  {Escher}}, \bibinfo {author} {\bibfnamefont {R.~L.}\ \bibnamefont
  {de~Matos~Filho}}, \ and\ \bibinfo {author} {\bibfnamefont {L.}~\bibnamefont
  {Davidovich}},\ }\href {https://doi.org/10.1038/nphys1958} {\bibfield
  {journal} {\bibinfo  {journal} {Nature Physics}\ }\textbf {\bibinfo {volume}
  {7}},\ \bibinfo {pages} {406} (\bibinfo {year} {2011})}\BibitemShut {NoStop}%
\bibitem [{\citenamefont {Demkowicz-Dobrza{\'{n}}ski}\ \emph
  {et~al.}(2012)\citenamefont {Demkowicz-Dobrza{\'{n}}ski}, \citenamefont
  {Ko{\l}ody{\'{n}}ski},\ and\ \citenamefont
  {Gu{\c{t}}{\u{a}}}}]{demkowicz2012elusive}%
  \BibitemOpen
  \bibfield  {author} {\bibinfo {author} {\bibfnamefont {R.}~\bibnamefont
  {Demkowicz-Dobrza{\'{n}}ski}}, \bibinfo {author} {\bibfnamefont
  {J.}~\bibnamefont {Ko{\l}ody{\'{n}}ski}}, \ and\ \bibinfo {author}
  {\bibfnamefont {M.}~\bibnamefont {Gu{\c{t}}{\u{a}}}},\ }\href
  {https://doi.org/10.1038/ncomms2067} {\bibfield  {journal} {\bibinfo
  {journal} {Nature Communications}\ }\textbf {\bibinfo {volume} {3}}, {\bibinfo {pages} {1063}} (\bibinfo
  {year} {2012})}\BibitemShut {NoStop}%
\bibitem [{\citenamefont {Leroux}\ \emph {et~al.}(2010)\citenamefont {Leroux},
  \citenamefont {Schleier-Smith},\ and\ \citenamefont
  {Vuleti{\'{c}}}}]{Leroux2010}%
  \BibitemOpen
  \bibfield  {author} {\bibinfo {author} {\bibfnamefont {I.~D.}\ \bibnamefont
  {Leroux}}, \bibinfo {author} {\bibfnamefont {M.~H.}\ \bibnamefont
  {Schleier-Smith}}, \ and\ \bibinfo {author} {\bibfnamefont {V.}~\bibnamefont
  {Vuleti{\'{c}}}},\ }\href {\doibase 10.1103/physrevlett.104.250801}
  {\bibfield  {journal} {\bibinfo  {journal} {Phys. Rev. Lett.}\
  }\textbf {\bibinfo {volume} {104}}, \bibinfo {pages} {250801}, (\bibinfo {year} {2010})}\
  \BibitemShut {NoStop}%
\bibitem [{\citenamefont {Appel}\ \emph {et~al.}(2009)\citenamefont {Appel},
  \citenamefont {Windpassinger}, \citenamefont {Oblak}, \citenamefont {Hoff},
  \citenamefont {Kjaergaard},\ and\ \citenamefont {Polzik}}]{Appel2009}%
  \BibitemOpen
  \bibfield  {author} {\bibinfo {author} {\bibfnamefont {J.}~\bibnamefont
  {Appel}}, \bibinfo {author} {\bibfnamefont {P.~J.}\ \bibnamefont
  {Windpassinger}}, \bibinfo {author} {\bibfnamefont {D.}~\bibnamefont
  {Oblak}}, \bibinfo {author} {\bibfnamefont {U.~B.}\ \bibnamefont {Hoff}},
  \bibinfo {author} {\bibfnamefont {N.}~\bibnamefont {Kjaergaard}}, \ and\
  \bibinfo {author} {\bibfnamefont {E.~S.}\ \bibnamefont {Polzik}},\ }\href
  {\doibase 10.1073/pnas.0901550106} {\bibfield  {journal} {\bibinfo  {journal}
  {Proceedings of the National Academy of Sciences}\ }\textbf {\bibinfo
  {volume} {106}},\ \bibinfo {pages} {10960} (\bibinfo {year}
  {2009})}\BibitemShut {NoStop}%
\bibitem [{\citenamefont {Gross}\ \emph {et~al.}(2010)\citenamefont {Gross},
  \citenamefont {Zibold}, \citenamefont {Nicklas}, \citenamefont
  {Est{\`{e}}ve},\ and\ \citenamefont {Oberthaler}}]{gross2012nonlinear}%
  \BibitemOpen
  \bibfield  {author} {\bibinfo {author} {\bibfnamefont {C.}~\bibnamefont
  {Gross}}, \bibinfo {author} {\bibfnamefont {T.}~\bibnamefont {Zibold}},
  \bibinfo {author} {\bibfnamefont {E.}~\bibnamefont {Nicklas}}, \bibinfo
  {author} {\bibfnamefont {J.}~\bibnamefont {Est{\`{e}}ve}}, \ and\ \bibinfo
  {author} {\bibfnamefont {M.~K.}\ \bibnamefont {Oberthaler}},\ }\href
  {https://doi.org/10.1038/nature08919} {\bibfield  {journal} {\bibinfo
  {journal} {Nature}\ }\textbf {\bibinfo {volume} {464}},\ \bibinfo {pages}
  {1165} (\bibinfo {year} {2010})}\BibitemShut {NoStop}%
\bibitem [{\citenamefont {Ockeloen}\ \emph {et~al.}(2013)\citenamefont
  {Ockeloen}, \citenamefont {Schmied}, \citenamefont {Riedel},\ and\
  \citenamefont {Treutlein}}]{ockeloen2013quantum}%
  \BibitemOpen
  \bibfield  {author} {\bibinfo {author} {\bibfnamefont {C.~F.}\ \bibnamefont
  {Ockeloen}}, \bibinfo {author} {\bibfnamefont {R.}~\bibnamefont {Schmied}},
  \bibinfo {author} {\bibfnamefont {M.~F.}\ \bibnamefont {Riedel}}, \ and\
  \bibinfo {author} {\bibfnamefont {P.}~\bibnamefont {Treutlein}},\ }\href
  {https://link.aps.org/doi/10.1103/PhysRevLett.111.143001} {\bibfield
  {journal} {\bibinfo  {journal} {Phys. Rev. Lett.}\ }\textbf {\bibinfo
  {volume} {111}},\ \bibinfo {pages} {143001} (\bibinfo {year}
  {2013})}\BibitemShut {NoStop}%
\bibitem [{\citenamefont {Muessel}\ \emph {et~al.}(2014)\citenamefont
  {Muessel}, \citenamefont {Strobel}, \citenamefont {Linnemann}, \citenamefont
  {Hume},\ and\ \citenamefont {Oberthaler}}]{mussel2014scalable}%
  \BibitemOpen
  \bibfield  {author} {\bibinfo {author} {\bibfnamefont {W.}~\bibnamefont
  {Muessel}}, \bibinfo {author} {\bibfnamefont {H.}~\bibnamefont {Strobel}},
  \bibinfo {author} {\bibfnamefont {D.}~\bibnamefont {Linnemann}}, \bibinfo
  {author} {\bibfnamefont {D.~B.}\ \bibnamefont {Hume}}, \ and\ \bibinfo
  {author} {\bibfnamefont {M.~K.}\ \bibnamefont {Oberthaler}},\ }\href
  {https://link.aps.org/doi/10.1103/PhysRevLett.113.103004} {\bibfield
  {journal} {\bibinfo  {journal} {Phys. Rev. Lett.}\ }\textbf {\bibinfo
  {volume} {113}},\ \bibinfo {pages} {103004} (\bibinfo {year}
  {2014})}\BibitemShut {NoStop}%
\bibitem [{\citenamefont {Strobel}\ \emph {et~al.}(2014)\citenamefont
  {Strobel}, \citenamefont {Muessel}, \citenamefont {Linnemann}, \citenamefont
  {Zibold}, \citenamefont {Hume}, \citenamefont {Pezze}, \citenamefont
  {Smerzi},\ and\ \citenamefont {Oberthaler}}]{strobel2014fisher}%
  \BibitemOpen
  \bibfield  {author} {\bibinfo {author} {\bibfnamefont {H.}~\bibnamefont
  {Strobel}}, \bibinfo {author} {\bibfnamefont {W.}~\bibnamefont {Muessel}},
  \bibinfo {author} {\bibfnamefont {D.}~\bibnamefont {Linnemann}}, \bibinfo
  {author} {\bibfnamefont {T.}~\bibnamefont {Zibold}}, \bibinfo {author}
  {\bibfnamefont {D.~B.}\ \bibnamefont {Hume}}, \bibinfo {author}
  {\bibfnamefont {L.}~\bibnamefont {Pezze}}, \bibinfo {author} {\bibfnamefont
  {A.}~\bibnamefont {Smerzi}}, \ and\ \bibinfo {author} {\bibfnamefont {M.~K.}\
  \bibnamefont {Oberthaler}},\ }\href {https://doi.org/10.1126/science.1250147}
  {\bibfield  {journal} {\bibinfo  {journal} {Science}\ }\textbf {\bibinfo
  {volume} {345}},\ \bibinfo {pages} {424} (\bibinfo {year}
  {2014})}\BibitemShut {NoStop}%
\bibitem [{\citenamefont {Lucke}\ \emph {et~al.}(2011)\citenamefont {Lucke},
  \citenamefont {Scherer}, \citenamefont {Kruse}, \citenamefont {Pezze},
  \citenamefont {Deuretzbacher}, \citenamefont {Hyllus}, \citenamefont {Topic},
  \citenamefont {Peise}, \citenamefont {Ertmer}, \citenamefont {Arlt},
  \citenamefont {Santos}, \citenamefont {Smerzi},\ and\ \citenamefont
  {Klempt}}]{lucke2011twin}%
  \BibitemOpen
  \bibfield  {author} {\bibinfo {author} {\bibfnamefont {B.}~\bibnamefont
  {Lucke}}, \bibinfo {author} {\bibfnamefont {M.}~\bibnamefont {Scherer}},
  \bibinfo {author} {\bibfnamefont {J.}~\bibnamefont {Kruse}}, \bibinfo
  {author} {\bibfnamefont {L.}~\bibnamefont {Pezze}}, \bibinfo {author}
  {\bibfnamefont {F.}~\bibnamefont {Deuretzbacher}}, \bibinfo {author}
  {\bibfnamefont {P.}~\bibnamefont {Hyllus}}, \bibinfo {author} {\bibfnamefont
  {O.}~\bibnamefont {Topic}}, \bibinfo {author} {\bibfnamefont
  {J.}~\bibnamefont {Peise}}, \bibinfo {author} {\bibfnamefont
  {W.}~\bibnamefont {Ertmer}}, \bibinfo {author} {\bibfnamefont
  {J.}~\bibnamefont {Arlt}}, \bibinfo {author} {\bibfnamefont {L.}~\bibnamefont
  {Santos}}, \bibinfo {author} {\bibfnamefont {A.}~\bibnamefont {Smerzi}}, \
  and\ \bibinfo {author} {\bibfnamefont {C.}~\bibnamefont {Klempt}},\ }\href
  {https://doi.org/10.1126/science.1208798} {\bibfield  {journal} {\bibinfo
  {journal} {Science}\ }\textbf {\bibinfo {volume} {334}},\ \bibinfo {pages}
  {773} (\bibinfo {year} {2011})}\BibitemShut {NoStop}%
\bibitem [{\citenamefont {Nagata}\ \emph {et~al.}(2007)\citenamefont {Nagata},
  \citenamefont {Okamoto}, \citenamefont {O'Brien}, \citenamefont {Sasaki},\
  and\ \citenamefont {Takeuchi}}]{nagata2007beating}%
  \BibitemOpen
  \bibfield  {author} {\bibinfo {author} {\bibfnamefont {T.}~\bibnamefont
  {Nagata}}, \bibinfo {author} {\bibfnamefont {R.}~\bibnamefont {Okamoto}},
  \bibinfo {author} {\bibfnamefont {J.~L.}\ \bibnamefont {O'Brien}}, \bibinfo
  {author} {\bibfnamefont {K.}~\bibnamefont {Sasaki}}, \ and\ \bibinfo {author}
  {\bibfnamefont {S.}~\bibnamefont {Takeuchi}},\ }\href
  {https://doi.org/10.1126/science.1138007} {\bibfield  {journal} {\bibinfo
  {journal} {Science}\ }\textbf {\bibinfo {volume} {316}},\ \bibinfo {pages}
  {726} (\bibinfo {year} {2007})}\BibitemShut {NoStop}%
\bibitem [{\citenamefont {Xiang}\ \emph {et~al.}(2010)\citenamefont {Xiang},
  \citenamefont {Higgins}, \citenamefont {Berry}, \citenamefont {Wiseman},\
  and\ \citenamefont {Pryde}}]{xiang2011entanglement}%
  \BibitemOpen
  \bibfield  {author} {\bibinfo {author} {\bibfnamefont {G.~Y.}\ \bibnamefont
  {Xiang}}, \bibinfo {author} {\bibfnamefont {B.~L.}\ \bibnamefont {Higgins}},
  \bibinfo {author} {\bibfnamefont {D.~W.}\ \bibnamefont {Berry}}, \bibinfo
  {author} {\bibfnamefont {H.~M.}\ \bibnamefont {Wiseman}}, \ and\ \bibinfo
  {author} {\bibfnamefont {G.~J.}\ \bibnamefont {Pryde}},\ }\href
  {https://doi.org/10.1038/nphoton.2010.268} {\bibfield  {journal} {\bibinfo
  {journal} {Nature Photonics}\ }\textbf {\bibinfo {volume} {5}},\ \bibinfo
  {pages} {43} (\bibinfo {year} {2010})}\BibitemShut {NoStop}%
\bibitem [{\citenamefont {Kacprowicz}\ \emph {et~al.}(2010)\citenamefont
  {Kacprowicz}, \citenamefont {Demkowicz-Dobrza{\'{n}}ski}, \citenamefont
  {Wasilewski}, \citenamefont {Banaszek},\ and\ \citenamefont
  {Walmsley}}]{kacprowicz2010experimental}%
  \BibitemOpen
  \bibfield  {author} {\bibinfo {author} {\bibfnamefont {M.}~\bibnamefont
  {Kacprowicz}}, \bibinfo {author} {\bibfnamefont {R.}~\bibnamefont
  {Demkowicz-Dobrza{\'{n}}ski}}, \bibinfo {author} {\bibfnamefont
  {W.}~\bibnamefont {Wasilewski}}, \bibinfo {author} {\bibfnamefont
  {K.}~\bibnamefont {Banaszek}}, \ and\ \bibinfo {author} {\bibfnamefont
  {I.~A.}\ \bibnamefont {Walmsley}},\ }\href
  {https://doi.org/10.1038/nphoton.2010.39} {\bibfield  {journal} {\bibinfo
  {journal} {Nature Photonics}\ }\textbf {\bibinfo {volume} {4}},\ \bibinfo
  {pages} {357} (\bibinfo {year} {2010})}\BibitemShut {NoStop}%
\bibitem [{\citenamefont {Afek}\ \emph {et~al.}(2010)\citenamefont {Afek},
  \citenamefont {Ambar},\ and\ \citenamefont {Silberberg}}]{afek2010high}%
  \BibitemOpen
  \bibfield  {author} {\bibinfo {author} {\bibfnamefont {I.}~\bibnamefont
  {Afek}}, \bibinfo {author} {\bibfnamefont {O.}~\bibnamefont {Ambar}}, \ and\
  \bibinfo {author} {\bibfnamefont {Y.}~\bibnamefont {Silberberg}},\ }\href
  {https://doi.org/10.1126/science.1188172} {\bibfield  {journal} {\bibinfo
  {journal} {Science}\ }\textbf {\bibinfo {volume} {328}},\ \bibinfo {pages}
  {879} (\bibinfo {year} {2010})}\BibitemShut {NoStop}%
\bibitem [{\citenamefont {Jachura}\ \emph {et~al.}(2016)\citenamefont
  {Jachura}, \citenamefont {Chrapkiewicz}, \citenamefont
  {Demkowicz-Dobrza{\'{n}}ski}, \citenamefont {Wasilewski},\ and\ \citenamefont
  {Banaszek}}]{jachura2016mode}%
  \BibitemOpen
  \bibfield  {author} {\bibinfo {author} {\bibfnamefont {M.}~\bibnamefont
  {Jachura}}, \bibinfo {author} {\bibfnamefont {R.}~\bibnamefont
  {Chrapkiewicz}}, \bibinfo {author} {\bibfnamefont {R.}~\bibnamefont
  {Demkowicz-Dobrza{\'{n}}ski}}, \bibinfo {author} {\bibfnamefont
  {W.}~\bibnamefont {Wasilewski}}, \ and\ \bibinfo {author} {\bibfnamefont
  {K.}~\bibnamefont {Banaszek}},\ }\href {https://doi.org/10.1038/ncomms11411}
  {\bibfield  {journal} {\bibinfo  {journal} {Nature Communications}\ }\textbf
  {\bibinfo {volume} {7}},\ \bibinfo {pages} {11411} (\bibinfo {year}
  {2016})}\BibitemShut {NoStop}%
\bibitem [{\citenamefont {Mivehvar}\ \emph {et~al.}(2018)\citenamefont
  {Mivehvar}, \citenamefont {Ostermann}, \citenamefont {Piazza},\ and\
  \citenamefont {Ritsch}}]{PhysRevLett.120.123601}%
  \BibitemOpen
  \bibfield  {author} {\bibinfo {author} {\bibfnamefont {F.}~\bibnamefont
  {Mivehvar}}, \bibinfo {author} {\bibfnamefont {S.}~\bibnamefont {Ostermann}},
  \bibinfo {author} {\bibfnamefont {F.}~\bibnamefont {Piazza}}, \ and\ \bibinfo
  {author} {\bibfnamefont {H.}~\bibnamefont {Ritsch}},\ }\href
  {https://link.aps.org/doi/10.1103/PhysRevLett.120.123601} {\bibfield
  {journal} {\bibinfo  {journal} {Phys. Rev. Lett.}\ }\textbf {\bibinfo
  {volume} {120}},\ \bibinfo {pages} {123601} (\bibinfo {year}
  {2018})}\BibitemShut {NoStop}%
\bibitem [{\citenamefont {Boninsegni}\ and\ \citenamefont
  {Prokof'ev}(2012)}]{RevModPhys.84.759}%
  \BibitemOpen
  \bibfield  {author} {\bibinfo {author} {\bibfnamefont {M.}~\bibnamefont
  {Boninsegni}}\ and\ \bibinfo {author} {\bibfnamefont {N.~V.}\ \bibnamefont
  {Prokof'ev}},\ }\href {https://link.aps.org/doi/10.1103/RevModPhys.84.759}
  {\bibfield  {journal} {\bibinfo  {journal} {Rev. Mod. Phys.}\ }\textbf
  {\bibinfo {volume} {84}},\ \bibinfo {pages} {759} (\bibinfo {year}
  {2012})}\BibitemShut {NoStop}%
\bibitem [{\citenamefont {Li}\ \emph {et~al.}(2017)\citenamefont {Li},
  \citenamefont {Lee}, \citenamefont {Huang}, \citenamefont {Burchesky},
  \citenamefont {Shteynas}, \citenamefont {Top}, \citenamefont {Jamison},\ and\
  \citenamefont {Ketterle}}]{li2017stripe}%
  \BibitemOpen
  \bibfield  {author} {\bibinfo {author} {\bibfnamefont {J.-R.}\ \bibnamefont
  {Li}}, \bibinfo {author} {\bibfnamefont {J.}~\bibnamefont {Lee}}, \bibinfo
  {author} {\bibfnamefont {W.}~\bibnamefont {Huang}}, \bibinfo {author}
  {\bibfnamefont {S.}~\bibnamefont {Burchesky}}, \bibinfo {author}
  {\bibfnamefont {B.}~\bibnamefont {Shteynas}}, \bibinfo {author}
  {\bibfnamefont {F.~{\c{C}}.}\ \bibnamefont {Top}}, \bibinfo {author}
  {\bibfnamefont {A.~O.}\ \bibnamefont {Jamison}}, \ and\ \bibinfo {author}
  {\bibfnamefont {W.}~\bibnamefont {Ketterle}},\ }\href
  {https://doi.org/10.1038/nature21431} {\bibfield  {journal} {\bibinfo
  {journal} {Nature}\ }\textbf {\bibinfo {volume} {543}},\ \bibinfo {pages}
  {91} (\bibinfo {year} {2017})}\BibitemShut {NoStop}%
\bibitem [{\citenamefont {L{\'{e}}onard}\ \emph {et~al.}(2017)\citenamefont
  {L{\'{e}}onard}, \citenamefont {Morales}, \citenamefont {Zupancic},
  \citenamefont {Esslinger},\ and\ \citenamefont
  {Donner}}]{leonard2017supersolid}%
  \BibitemOpen
  \bibfield  {author} {\bibinfo {author} {\bibfnamefont {J.}~\bibnamefont
  {L{\'{e}}onard}}, \bibinfo {author} {\bibfnamefont {A.}~\bibnamefont
  {Morales}}, \bibinfo {author} {\bibfnamefont {P.}~\bibnamefont {Zupancic}},
  \bibinfo {author} {\bibfnamefont {T.}~\bibnamefont {Esslinger}}, \ and\
  \bibinfo {author} {\bibfnamefont {T.}~\bibnamefont {Donner}},\ }\href
  {https://doi.org/10.1038/nature21067} {\bibfield  {journal} {\bibinfo
  {journal} {Nature}\ }\textbf {\bibinfo {volume} {543}},\ \bibinfo {pages}
  {87} (\bibinfo {year} {2017})}\BibitemShut {NoStop}%
\bibitem [{\citenamefont {Maschler}\ \emph {et~al.}(2007)\citenamefont
  {Maschler}, \citenamefont {Ritsch}, \citenamefont {Vukics},\ and\
  \citenamefont {Domokos}}]{maschler2007entanglement}%
  \BibitemOpen
  \bibfield  {author} {\bibinfo {author} {\bibfnamefont {C.}~\bibnamefont
  {Maschler}}, \bibinfo {author} {\bibfnamefont {H.}~\bibnamefont {Ritsch}},
  \bibinfo {author} {\bibfnamefont {A.}~\bibnamefont {Vukics}}, \ and\ \bibinfo
  {author} {\bibfnamefont {P.}~\bibnamefont {Domokos}},\ }\href
  {https://doi.org/10.1016/j.optcom.2007.01.069} {\bibfield  {journal}
  {\bibinfo  {journal} {Optics Communications}\ }\textbf {\bibinfo {volume}
  {273}},\ \bibinfo {pages} {446} (\bibinfo {year} {2007})}\BibitemShut
  {NoStop}%
\bibitem [{\citenamefont {Kr\"amer}\ and\ \citenamefont
  {Ritsch}(2014)}]{kramer2014self}%
  \BibitemOpen
  \bibfield  {author} {\bibinfo {author} {\bibfnamefont {S.}~\bibnamefont
  {Kr\"amer}}\ and\ \bibinfo {author} {\bibfnamefont {H.}~\bibnamefont
  {Ritsch}},\ }\href {https://link.aps.org/doi/10.1103/PhysRevA.90.033833}
  {\bibfield  {journal} {\bibinfo  {journal} {Phys. Rev. A}\ }\textbf {\bibinfo
  {volume} {90}},\ \bibinfo {pages} {033833} (\bibinfo {year}
  {2014})}\BibitemShut {NoStop}%
\bibitem [{\citenamefont {Vukics}\ \emph {et~al.}(2007)\citenamefont {Vukics},
  \citenamefont {Maschler},\ and\ \citenamefont
  {Ritsch}}]{vukics2007microscopic}%
  \BibitemOpen
  \bibfield  {author} {\bibinfo {author} {\bibfnamefont {A.}~\bibnamefont
  {Vukics}}, \bibinfo {author} {\bibfnamefont {C.}~\bibnamefont {Maschler}}, \
  and\ \bibinfo {author} {\bibfnamefont {H.}~\bibnamefont {Ritsch}},\ }\href
  {https://doi.org/10.1088/1367-2630/9/8/255} {\bibfield  {journal} {\bibinfo
  {journal} {New J. Phys.}\ }\textbf {\bibinfo {volume} {9}},\ \bibinfo {pages}
  {255} (\bibinfo {year} {2007})}\BibitemShut {NoStop}%
\bibitem [{SM-()}]{SM-SS-gravimeter-2018}%
  \BibitemOpen
  \href@noop {} {\bibinfo  {journal} {See Supplemental Material for details and
  some results of the full quantum-mechanical calculations.}\ }\BibitemShut
  {NoStop}%
\bibitem [{\citenamefont {Kruse}\ \emph
  {et~al.}(2003{\natexlab{a}})\citenamefont {Kruse}, \citenamefont {von Cube},
  \citenamefont {Zimmermann},\ and\ \citenamefont
  {Courteille}}]{Kruse2003atomic-rcoil-laser}%
  \BibitemOpen
\bibfield  {journal} {  }\bibfield  {author} {\bibinfo {author} {\bibfnamefont
  {D.}~\bibnamefont {Kruse}}, \bibinfo {author} {\bibfnamefont
  {C.}~\bibnamefont {von Cube}}, \bibinfo {author} {\bibfnamefont
  {C.}~\bibnamefont {Zimmermann}}, \ and\ \bibinfo {author} {\bibfnamefont
  {P.~W.}\ \bibnamefont {Courteille}},\ }\href
  {https://link.aps.org/doi/10.1103/PhysRevLett.91.183601} {\bibfield
  {journal} {\bibinfo  {journal} {Phys. Rev. Lett.}\ }\textbf {\bibinfo
  {volume} {91}},\ \bibinfo {pages} {183601} (\bibinfo {year}
  {2003}{\natexlab{a}})}\BibitemShut {NoStop}%
\bibitem [{\citenamefont {Kruse}\ \emph
  {et~al.}(2003{\natexlab{b}})\citenamefont {Kruse}, \citenamefont {Ruder},
  \citenamefont {Benhelm}, \citenamefont {von Cube}, \citenamefont
  {Zimmermann}, \citenamefont {Courteille}, \citenamefont {Els\"asser},
  \citenamefont {Nagorny},\ and\ \citenamefont {Hemmerich}}]{kruse2003cold}%
  \BibitemOpen
  \bibfield  {author} {\bibinfo {author} {\bibfnamefont {D.}~\bibnamefont
  {Kruse}}, \bibinfo {author} {\bibfnamefont {M.}~\bibnamefont {Ruder}},
  \bibinfo {author} {\bibfnamefont {J.}~\bibnamefont {Benhelm}}, \bibinfo
  {author} {\bibfnamefont {C.}~\bibnamefont {von Cube}}, \bibinfo {author}
  {\bibfnamefont {C.}~\bibnamefont {Zimmermann}}, \bibinfo {author}
  {\bibfnamefont {P.~W.}\ \bibnamefont {Courteille}}, \bibinfo {author}
  {\bibfnamefont {T.}~\bibnamefont {Els\"asser}}, \bibinfo {author}
  {\bibfnamefont {B.}~\bibnamefont {Nagorny}}, \ and\ \bibinfo {author}
  {\bibfnamefont {A.}~\bibnamefont {Hemmerich}},\ }\href
  {https://link.aps.org/doi/10.1103/PhysRevA.67.051802} {\bibfield  {journal}
  {\bibinfo  {journal} {Phys. Rev. A}\ }\textbf {\bibinfo {volume} {67}},\
  \bibinfo {pages} {051802(R)} (\bibinfo {year} {2003}{\natexlab{b}})}\BibitemShut
  {NoStop}%
\bibitem [{\citenamefont {Nagorny}\ \emph {et~al.}(2003)\citenamefont
  {Nagorny}, \citenamefont {Els\"asser}, \citenamefont {Richter}, \citenamefont
  {Hemmerich}, \citenamefont {Kruse}, \citenamefont {Zimmermann},\ and\
  \citenamefont {Courteille}}]{nagorny2003optical}%
  \BibitemOpen
  \bibfield  {author} {\bibinfo {author} {\bibfnamefont {B.}~\bibnamefont
  {Nagorny}}, \bibinfo {author} {\bibfnamefont {T.}~\bibnamefont {Els\"asser}},
  \bibinfo {author} {\bibfnamefont {H.}~\bibnamefont {Richter}}, \bibinfo
  {author} {\bibfnamefont {A.}~\bibnamefont {Hemmerich}}, \bibinfo {author}
  {\bibfnamefont {D.}~\bibnamefont {Kruse}}, \bibinfo {author} {\bibfnamefont
  {C.}~\bibnamefont {Zimmermann}}, \ and\ \bibinfo {author} {\bibfnamefont
  {P.}~\bibnamefont {Courteille}},\ }\href
  {https://link.aps.org/doi/10.1103/PhysRevA.67.031401} {\bibfield  {journal}
  {\bibinfo  {journal} {Phys. Rev. A}\ }\textbf {\bibinfo {volume} {67}},\
  \bibinfo {pages} {031401(R)} (\bibinfo {year} {2003})}\BibitemShut {NoStop}%
\bibitem [{\citenamefont {Slama}\ \emph
  {et~al.}(2007{\natexlab{a}})\citenamefont {Slama}, \citenamefont {Bux},
  \citenamefont {Krenz}, \citenamefont {Zimmermann},\ and\ \citenamefont
  {Courteille}}]{Slama2007Rayleigh-scattering}%
  \BibitemOpen
  \bibfield  {author} {\bibinfo {author} {\bibfnamefont {S.}~\bibnamefont
  {Slama}}, \bibinfo {author} {\bibfnamefont {S.}~\bibnamefont {Bux}}, \bibinfo
  {author} {\bibfnamefont {G.}~\bibnamefont {Krenz}}, \bibinfo {author}
  {\bibfnamefont {C.}~\bibnamefont {Zimmermann}}, \ and\ \bibinfo {author}
  {\bibfnamefont {P.~W.}\ \bibnamefont {Courteille}},\ }\href {\doibase
  10.1103/PhysRevLett.98.053603} {\bibfield  {journal} {\bibinfo  {journal}
  {Phys. Rev. Lett.}\ }\textbf {\bibinfo {volume} {98}},\ \bibinfo {pages}
  {053603} (\bibinfo {year} {2007}{\natexlab{a}})}\BibitemShut {NoStop}%
\bibitem [{\citenamefont {Slama}\ \emph
  {et~al.}(2007{\natexlab{b}})\citenamefont {Slama}, \citenamefont {Krenz},
  \citenamefont {Bux}, \citenamefont {Zimmermann},\ and\ \citenamefont
  {Courteille}}]{slama2007cavity}%
  \BibitemOpen
  \bibfield  {author} {\bibinfo {author} {\bibfnamefont {S.}~\bibnamefont
  {Slama}}, \bibinfo {author} {\bibfnamefont {G.}~\bibnamefont {Krenz}},
  \bibinfo {author} {\bibfnamefont {S.}~\bibnamefont {Bux}}, \bibinfo {author}
  {\bibfnamefont {C.}~\bibnamefont {Zimmermann}}, \ and\ \bibinfo {author}
  {\bibfnamefont {P.~W.}\ \bibnamefont {Courteille}},\ }\href
  {https://link.aps.org/doi/10.1103/PhysRevA.75.063620} {\bibfield  {journal}
  {\bibinfo  {journal} {Phys. Rev. A}\ }\textbf {\bibinfo {volume} {75}},\
  \bibinfo {pages} {063620} (\bibinfo {year} {2007}{\natexlab{b}})}\BibitemShut
  {NoStop}%
\bibitem [{\citenamefont {Bux}\ \emph {et~al.}(2013)\citenamefont {Bux},
  \citenamefont {Tomczyk}, \citenamefont {Schmidt}, \citenamefont {Courteille},
  \citenamefont {Piovella},\ and\ \citenamefont {Zimmermann}}]{bux2013control}%
  \BibitemOpen
  \bibfield  {author} {\bibinfo {author} {\bibfnamefont {S.}~\bibnamefont
  {Bux}}, \bibinfo {author} {\bibfnamefont {H.}~\bibnamefont {Tomczyk}},
  \bibinfo {author} {\bibfnamefont {D.}~\bibnamefont {Schmidt}}, \bibinfo
  {author} {\bibfnamefont {P.~W.}\ \bibnamefont {Courteille}}, \bibinfo
  {author} {\bibfnamefont {N.}~\bibnamefont {Piovella}}, \ and\ \bibinfo
  {author} {\bibfnamefont {C.}~\bibnamefont {Zimmermann}},\ }\href
  {https://link.aps.org/doi/10.1103/PhysRevA.87.023607} {\bibfield  {journal}
  {\bibinfo  {journal} {Phys. Rev. A}\ }\textbf {\bibinfo {volume} {87}},\
  \bibinfo {pages} {023607} (\bibinfo {year} {2013})}\BibitemShut {NoStop}%
\bibitem [{\citenamefont {Schmidt}\ \emph {et~al.}(2014)\citenamefont
  {Schmidt}, \citenamefont {Tomczyk}, \citenamefont {Slama},\ and\
  \citenamefont {Zimmermann}}]{schmidt2014dynamical}%
  \BibitemOpen
  \bibfield  {author} {\bibinfo {author} {\bibfnamefont {D.}~\bibnamefont
  {Schmidt}}, \bibinfo {author} {\bibfnamefont {H.}~\bibnamefont {Tomczyk}},
  \bibinfo {author} {\bibfnamefont {S.}~\bibnamefont {Slama}}, \ and\ \bibinfo
  {author} {\bibfnamefont {C.}~\bibnamefont {Zimmermann}},\ }\href
  {https://link.aps.org/doi/10.1103/PhysRevLett.112.115302} {\bibfield
  {journal} {\bibinfo  {journal} {Phys. Rev. Lett.}\ }\textbf {\bibinfo
  {volume} {112}},\ \bibinfo {pages} {115302} (\bibinfo {year}
  {2014})}\BibitemShut {NoStop}%
\bibitem [{\citenamefont {Culver}\ \emph {et~al.}(2016)\citenamefont {Culver},
  \citenamefont {Lampis}, \citenamefont {Megyeri}, \citenamefont {Pahwa},
  \citenamefont {Mudarikwa}, \citenamefont {Holynski}, \citenamefont
  {Courteille},\ and\ \citenamefont {Goldwin}}]{Culver2016}%
  \BibitemOpen
  \bibfield  {author} {\bibinfo {author} {\bibfnamefont {R.}~\bibnamefont
  {Culver}}, \bibinfo {author} {\bibfnamefont {A.}~\bibnamefont {Lampis}},
  \bibinfo {author} {\bibfnamefont {B.}~\bibnamefont {Megyeri}}, \bibinfo
  {author} {\bibfnamefont {K.}~\bibnamefont {Pahwa}}, \bibinfo {author}
  {\bibfnamefont {L.}~\bibnamefont {Mudarikwa}}, \bibinfo {author}
  {\bibfnamefont {M.}~\bibnamefont {Holynski}}, \bibinfo {author}
  {\bibfnamefont {P.~W.}\ \bibnamefont {Courteille}}, \ and\ \bibinfo {author}
  {\bibfnamefont {J.}~\bibnamefont {Goldwin}},\ }\href
  {https://doi.org/10.1088/1367-2630/18/11/113043} {\bibfield  {journal}
  {\bibinfo  {journal} {New J. Phys.}\ }\textbf {\bibinfo {volume} {18}},\
  \bibinfo {pages} {113043} (\bibinfo {year} {2016})}\BibitemShut {NoStop}%
\bibitem [{\citenamefont {Naik}\ \emph {et~al.}(2018)\citenamefont {Naik},
  \citenamefont {Kuyumjyan}, \citenamefont {Pandey}, \citenamefont {Bouyer},\
  and\ \citenamefont {Bertoldi}}]{Naik2018}%
  \BibitemOpen
  \bibfield  {author} {\bibinfo {author} {\bibfnamefont {D.~S.}\ \bibnamefont
  {Naik}}, \bibinfo {author} {\bibfnamefont {G.}~\bibnamefont {Kuyumjyan}},
  \bibinfo {author} {\bibfnamefont {D.}~\bibnamefont {Pandey}}, \bibinfo
  {author} {\bibfnamefont {P.}~\bibnamefont {Bouyer}}, \ and\ \bibinfo {author}
  {\bibfnamefont {A.}~\bibnamefont {Bertoldi}},\ }\href
  {https://doi.org/10.1088/2058-9565/aad48e} {\bibfield  {journal} {\bibinfo
  {journal} {Quantum Science and Technology}\ }\textbf {\bibinfo {volume}
  {3}},\ \bibinfo {pages} {045009} (\bibinfo {year} {2018})}\BibitemShut
  {NoStop}%
\bibitem [{\citenamefont {Cox}\ \emph {et~al.}(2018)\citenamefont {Cox},
  \citenamefont {Meyer}, \citenamefont {Schine}, \citenamefont {Fatemi},\ and\
  \citenamefont {Kunz}}]{Cox2018ringcavity}%
  \BibitemOpen
  \bibfield  {author} {\bibinfo {author} {\bibfnamefont {K.~C.}\ \bibnamefont
  {Cox}}, \bibinfo {author} {\bibfnamefont {D.~H.}\ \bibnamefont {Meyer}},
  \bibinfo {author} {\bibfnamefont {N.~A.}\ \bibnamefont {Schine}}, \bibinfo
  {author} {\bibfnamefont {F.~K.}\ \bibnamefont {Fatemi}}, \ and\ \bibinfo
  {author} {\bibfnamefont {P.~D.}\ \bibnamefont {Kunz}},\ }\href {\doibase
  10.1088/1361-6455/aaddd1} {\bibfield  {journal} {\bibinfo  {journal} {Journal
  of Physics B: Atomic, Molecular and Optical Physics}\ }\textbf {\bibinfo
  {volume} {51}},\ \bibinfo {pages} {195002} (\bibinfo {year}
  {2018})}\BibitemShut {NoStop}%
\bibitem [{\citenamefont {Wolf}\ \emph {et~al.}(2018)\citenamefont {Wolf},
  \citenamefont {Schuster}, \citenamefont {Schmidt}, \citenamefont {Slama},\
  and\ \citenamefont {Zimmermann}}]{Wolf2018}%
  \BibitemOpen
  \bibfield  {author} {\bibinfo {author} {\bibfnamefont {P.}~\bibnamefont
  {Wolf}}, \bibinfo {author} {\bibfnamefont {S.C.}~\bibnamefont {Schuster}},
  \bibinfo {author} {\bibfnamefont {D.}~\bibnamefont {Schmidt}}, \bibinfo
  {author} {\bibfnamefont {S.}~\bibnamefont {Slama}}, \ and\ \bibinfo {author}
  {\bibfnamefont {C.}~\bibnamefont {Zimmermann}},\ }\href
  {https://doi.org/10.1103/physrevlett.121.173602} {\bibfield  {journal}
  {\bibinfo  {journal} {Phys. Rev. Lett.}\ }\textbf {\bibinfo {volume} {121}}, {\bibinfo {pages} {173602}}
  (\bibinfo {year} {2018})}\BibitemShut {NoStop}%
\bibitem [{\citenamefont {Schuster}\ \emph {et~al.}(2018)\citenamefont
  {Schuster}, \citenamefont {Wolf}, \citenamefont {Schmidt}, \citenamefont
  {Slama},\ and\ \citenamefont {Zimmermann}}]{schuster2018pinning}%
  \BibitemOpen
  \bibfield  {author} {\bibinfo {author} {\bibfnamefont {S.C.}~\bibnamefont
  {Schuster}}, \bibinfo {author} {\bibfnamefont {P.}~\bibnamefont {Wolf}},
  \bibinfo {author} {\bibfnamefont {D.}~\bibnamefont {Schmidt}}, \bibinfo
  {author} {\bibfnamefont {S.}~\bibnamefont {Slama}}, \ and\ \bibinfo {author}
  {\bibfnamefont {C.}~\bibnamefont {Zimmermann}},\ }\href
  {https://doi.org/10.1103/PhysRevLett.121.223601} {\bibfield  {journal} {\bibinfo
  {journal} {Phys. Rev. Lett.}\ } \textbf {\bibinfo {volume} {121}}, {\bibinfo {pages} {223601}} (\bibinfo {year} {2018})}\BibitemShut
  {NoStop}%
\bibitem [{\citenamefont {Piazza}\ \emph {et~al.}(2013)\citenamefont {Piazza},
  \citenamefont {Strack},\ and\ \citenamefont {Zwerger}}]{piazza2013bose}%
  \BibitemOpen
  \bibfield  {author} {\bibinfo {author} {\bibfnamefont {F.}~\bibnamefont
  {Piazza}}, \bibinfo {author} {\bibfnamefont {P.}~\bibnamefont {Strack}}, \
  and\ \bibinfo {author} {\bibfnamefont {W.}~\bibnamefont {Zwerger}},\ }\href
  {\doibase 10.1016/j.aop.2013.08.015} {\bibfield  {journal} {\bibinfo
  {journal} {Annals of Physics}\ }\textbf {\bibinfo {volume} {339}},\ \bibinfo
  {pages} {135} (\bibinfo {year} {2013})}\BibitemShut {NoStop}%
\bibitem [{Note1()}]{Note1}%
  \BibitemOpen
  \bibinfo {note} {The modification of the cavity-field populations due to the
  presence of the gravitational potential are negligible for deep cavity
  potentials. That is, the cavity-field populations are almost the same as
  steady-state mean photon numbers in the absence of the gravitational
  potential}\BibitemShut {NoStop}%
\bibitem [{\citenamefont {Gangl}\ and\ \citenamefont
  {Ritsch}(2000)}]{gangl2000cold}%
  \BibitemOpen
  \bibfield  {author} {\bibinfo {author} {\bibfnamefont {M.}~\bibnamefont
  {Gangl}}\ and\ \bibinfo {author} {\bibfnamefont {H.}~\bibnamefont {Ritsch}},\
  }\href {https://doi.org/10.1103/physreva.61.043405} {\bibfield  {journal}
  {\bibinfo  {journal} {Phys. Rev. A}\ }\textbf {\bibinfo {volume} {61}}, {\bibinfo {pages} {043405}}
  (\bibinfo {year} {2000})}\BibitemShut {NoStop}%
\bibitem [{\citenamefont {Ostermann}\ \emph {et~al.}(2013)\citenamefont
  {Ostermann}, \citenamefont {Ritsch},\ and\ \citenamefont
  {Genes}}]{PhysRevLett.111.123601}%
  \BibitemOpen
  \bibfield  {author} {\bibinfo {author} {\bibfnamefont {L.}~\bibnamefont
  {Ostermann}}, \bibinfo {author} {\bibfnamefont {H.}~\bibnamefont {Ritsch}}, \
  and\ \bibinfo {author} {\bibfnamefont {C.}~\bibnamefont {Genes}},\ }\href
  {https://link.aps.org/doi/10.1103/PhysRevLett.111.123601} {\bibfield
  {journal} {\bibinfo  {journal} {Phys. Rev. Lett.}\ }\textbf {\bibinfo
  {volume} {111}},\ \bibinfo {pages} {123601} (\bibinfo {year}
  {2013})}\BibitemShut {NoStop}%
\bibitem [{\citenamefont {Kasevich}\ and\ \citenamefont
  {Chu}(1991)}]{kasevich1991atomic}%
  \BibitemOpen
  \bibfield  {author} {\bibinfo {author} {\bibfnamefont {M.}~\bibnamefont
  {Kasevich}}\ and\ \bibinfo {author} {\bibfnamefont {S.}~\bibnamefont {Chu}},\
  }\href {https://doi.org/10.1103/physrevlett.67.181} {\bibfield  {journal}
  {\bibinfo  {journal} {Phys. Rev. Lett.}\ }\textbf {\bibinfo {volume} {67}},\
  \bibinfo {pages} {181} (\bibinfo {year} {1991})}\BibitemShut {NoStop}%
\bibitem [{\citenamefont {Bord{\'{e}}}(1989)}]{borde1989atomic}%
  \BibitemOpen
  \bibfield  {author} {\bibinfo {author} {\bibfnamefont {C.}~\bibnamefont
  {Bord{\'{e}}}},\ }\href {https://doi.org/10.1016/0375-9601(89)90537-9}
  {\bibfield  {journal} {\bibinfo  {journal} {Physics Letters A}\ }\textbf
  {\bibinfo {volume} {140}},\ \bibinfo {pages} {10} (\bibinfo {year}
  {1989})}\BibitemShut {NoStop}%
\bibitem [{\citenamefont {Kasevich}\ and\ \citenamefont
  {Chu}(1992)}]{kasevich1992measurement}%
  \BibitemOpen
  \bibfield  {author} {\bibinfo {author} {\bibfnamefont {M.}~\bibnamefont
  {Kasevich}}\ and\ \bibinfo {author} {\bibfnamefont {S.}~\bibnamefont {Chu}},\
  }\href {https://doi.org/10.1007/bf00325375} {\bibfield  {journal} {\bibinfo
  {journal} {Applied Physics B Photophysics and Laser Chemistry}\ }\textbf
  {\bibinfo {volume} {54}},\ \bibinfo {pages} {321} (\bibinfo {year}
  {1992})}\BibitemShut {NoStop}%
\bibitem [{\citenamefont {Schleich}\ \emph {et~al.}(2013)\citenamefont
  {Schleich}, \citenamefont {Greenberger},\ and\ \citenamefont
  {Rasel}}]{schleich2013redshift}%
  \BibitemOpen
  \bibfield  {author} {\bibinfo {author} {\bibfnamefont {W.~P.}\ \bibnamefont
  {Schleich}}, \bibinfo {author} {\bibfnamefont {D.~M.}\ \bibnamefont
  {Greenberger}}, \ and\ \bibinfo {author} {\bibfnamefont {E.~M.}\ \bibnamefont
  {Rasel}},\ }\href {https://doi.org/10.1103/physrevlett.110.010401} {\bibfield
   {journal} {\bibinfo  {journal} {Phys. Rev. Lett.}\ }\textbf {\bibinfo
  {volume} {110}}, {\bibinfo {pages} {010401}} (\bibinfo {year} {2013})}\BibitemShut {NoStop}%
\bibitem [{\citenamefont {Braun}\ \emph {et~al.}(2018)\citenamefont {Braun},
  \citenamefont {Adesso}, \citenamefont {Benatti}, \citenamefont {Floreanini},
  \citenamefont {Marzolino}, \citenamefont {Mitchell},\ and\ \citenamefont
  {Pirandola}}]{RevModPhys.90.035006}%
  \BibitemOpen
  \bibfield  {author} {\bibinfo {author} {\bibfnamefont {D.}~\bibnamefont
  {Braun}}, \bibinfo {author} {\bibfnamefont {G.}~\bibnamefont {Adesso}},
  \bibinfo {author} {\bibfnamefont {F.}~\bibnamefont {Benatti}}, \bibinfo
  {author} {\bibfnamefont {R.}~\bibnamefont {Floreanini}}, \bibinfo {author}
  {\bibfnamefont {U.}~\bibnamefont {Marzolino}}, \bibinfo {author}
  {\bibfnamefont {M.~W.}\ \bibnamefont {Mitchell}}, \ and\ \bibinfo {author}
  {\bibfnamefont {S.}~\bibnamefont {Pirandola}},\ }\href
  {https://link.aps.org/doi/10.1103/RevModPhys.90.035006} {\bibfield  {journal}
  {\bibinfo  {journal} {Rev. Mod. Phys.}\ }\textbf {\bibinfo {volume} {90}},\
  \bibinfo {pages} {035006} (\bibinfo {year} {2018})}\BibitemShut {NoStop}%
\bibitem [{\citenamefont {Fisher}(1925)}]{Fisher1925}%
  \BibitemOpen
  \bibfield  {author} {\bibinfo {author} {\bibfnamefont {R.~A.}\ \bibnamefont
  {Fisher}},\ }\href {https://doi.org/10.1017/s0305004100009580} {\bibfield
  {journal} {\bibinfo  {journal} {Mathematical Proceedings of the Cambridge
  Philosophical Society}\ }\textbf {\bibinfo {volume} {22}},\ \bibinfo {pages}
  {700} (\bibinfo {year} {1925})}\BibitemShut {NoStop}%
\bibitem [{\citenamefont {Braunstein}\ and\ \citenamefont
  {Caves}(1994)}]{Braunstein1994}%
  \BibitemOpen
  \bibfield  {author} {\bibinfo {author} {\bibfnamefont {S.~L.}\ \bibnamefont
  {Braunstein}}\ and\ \bibinfo {author} {\bibfnamefont {C.~M.}\ \bibnamefont
  {Caves}},\ }\href {https://doi.org/10.1103/physrevlett.72.3439} {\bibfield
  {journal} {\bibinfo  {journal} {Phys. Rev. Lett.}\ }\textbf {\bibinfo
  {volume} {72}},\ \bibinfo {pages} {3439} (\bibinfo {year}
  {1994})}\BibitemShut {NoStop}%
\bibitem [{\citenamefont {Pinel}\ \emph {et~al.}(2013)\citenamefont {Pinel},
  \citenamefont {Jian}, \citenamefont {Treps}, \citenamefont {Fabre},\ and\
  \citenamefont {Braun}}]{pinel2013quantum}%
  \BibitemOpen
  \bibfield  {author} {\bibinfo {author} {\bibfnamefont {O.}~\bibnamefont
  {Pinel}}, \bibinfo {author} {\bibfnamefont {P.}~\bibnamefont {Jian}},
  \bibinfo {author} {\bibfnamefont {N.}~\bibnamefont {Treps}}, \bibinfo
  {author} {\bibfnamefont {C.}~\bibnamefont {Fabre}}, \ and\ \bibinfo {author}
  {\bibfnamefont {D.}~\bibnamefont {Braun}},\ }\href
  {https://link.aps.org/doi/10.1103/PhysRevA.88.040102} {\bibfield  {journal}
  {\bibinfo  {journal} {Phys. Rev. A}\ }\textbf {\bibinfo {volume} {88}},\
  \bibinfo {pages} {040102(R)} (\bibinfo {year} {2013})}\BibitemShut {NoStop}%
\bibitem [{\citenamefont {Pezz{\'{e}}}\ and\ \citenamefont
  {Smerzi}(2008)}]{Pezz2008}%
  \BibitemOpen
  \bibfield  {author} {\bibinfo {author} {\bibfnamefont {L.}~\bibnamefont
  {Pezz{\'{e}}}}\ and\ \bibinfo {author} {\bibfnamefont {A.}~\bibnamefont
  {Smerzi}},\ }\href {https://doi.org/10.1103/physrevlett.100.073601}
  {\bibfield  {journal} {\bibinfo  {journal} {Phys. Rev. Lett.}\
  }\textbf {\bibinfo {volume} {100}}, {\bibinfo {pages} {073601}} (\bibinfo {year} {2008})}\BibitemShut
  {NoStop}%
\bibitem [{\citenamefont {Cervantes}\ \emph {et~al.}(2014)\citenamefont
  {Cervantes}, \citenamefont {Kumanchik}, \citenamefont {Pratt},\ and\
  \citenamefont {Taylor}}]{guzman2014high}%
  \BibitemOpen
  \bibfield  {author} {\bibinfo {author} {\bibfnamefont {F.~G.}\ \bibnamefont
  {Cervantes}}, \bibinfo {author} {\bibfnamefont {L.}~\bibnamefont
  {Kumanchik}}, \bibinfo {author} {\bibfnamefont {J.}~\bibnamefont {Pratt}}, \
  and\ \bibinfo {author} {\bibfnamefont {J.~M.}\ \bibnamefont {Taylor}},\
  }\href {https://doi.org/10.1063/1.4881936} {\bibfield  {journal} {\bibinfo
  {journal} {Applied Physics Letters}\ }\textbf {\bibinfo {volume} {104}},\
  \bibinfo {pages} {221111} (\bibinfo {year} {2014})}\BibitemShut {NoStop}%
\bibitem [{lac()}]{lacoste}%
  \BibitemOpen
  \href@noop {} {}\bibinfo {howpublished} {\url{http://microglacoste.com}},\
  \bibinfo {note} {micro-g LaCoste, I. FG5-X Absolute Gravimeter}\BibitemShut
  {NoStop}%
\bibitem [{\citenamefont {Hardman}\ \emph {et~al.}(2016)\citenamefont
  {Hardman}, \citenamefont {Everitt}, \citenamefont {McDonald}, \citenamefont
  {Manju}, \citenamefont {Wigley}, \citenamefont {Sooriyabandara},
  \citenamefont {Kuhn}, \citenamefont {Debs}, \citenamefont {Close},\ and\
  \citenamefont {Robins}}]{hardman2016simultaneous}%
  \BibitemOpen
  \bibfield  {author} {\bibinfo {author} {\bibfnamefont {K.~S.}\ \bibnamefont
  {Hardman}}, \bibinfo {author} {\bibfnamefont {P.~J.}\ \bibnamefont
  {Everitt}}, \bibinfo {author} {\bibfnamefont {G.~D.}\ \bibnamefont
  {McDonald}}, \bibinfo {author} {\bibfnamefont {P.}~\bibnamefont {Manju}},
  \bibinfo {author} {\bibfnamefont {P.~B.}\ \bibnamefont {Wigley}}, \bibinfo
  {author} {\bibfnamefont {M.~A.}\ \bibnamefont {Sooriyabandara}}, \bibinfo
  {author} {\bibfnamefont {C.~C.~N.}\ \bibnamefont {Kuhn}}, \bibinfo {author}
  {\bibfnamefont {J.~E.}\ \bibnamefont {Debs}}, \bibinfo {author}
  {\bibfnamefont {J.~D.}\ \bibnamefont {Close}}, \ and\ \bibinfo {author}
  {\bibfnamefont {N.~P.}\ \bibnamefont {Robins}},\ }\href
  {https://link.aps.org/doi/10.1103/PhysRevLett.117.138501} {\bibfield
  {journal} {\bibinfo  {journal} {Phys. Rev. Lett.}\ }\textbf {\bibinfo
  {volume} {117}},\ \bibinfo {pages} {138501} (\bibinfo {year}
  {2016})}\BibitemShut {NoStop}%
\bibitem [{\citenamefont {Kr\"{a}mer}\ \emph {et~al.}(2018)\citenamefont
  {Kr\"{a}mer}, \citenamefont {Plankensteiner}, \citenamefont {Ostermann},\
  and\ \citenamefont {Ritsch}}]{kramer2018quantumoptics}%
  \BibitemOpen
  \bibfield  {author} {\bibinfo {author} {\bibfnamefont {S.}~\bibnamefont
  {Kr\"{a}mer}}, \bibinfo {author} {\bibfnamefont {D.}~\bibnamefont
  {Plankensteiner}}, \bibinfo {author} {\bibfnamefont {L.}~\bibnamefont
  {Ostermann}}, \ and\ \bibinfo {author} {\bibfnamefont {H.}~\bibnamefont
  {Ritsch}},\ }\href {https://doi.org/10.1016/j.cpc.2018.02.004} {\bibfield
  {journal} {\bibinfo  {journal} {Computer Physics Communications}\ }\textbf
  {\bibinfo {volume} {227}},\ \bibinfo {pages} {109} (\bibinfo {year}
  {2018})}\BibitemShut {NoStop}%
\end{thebibliography}
%

\onecolumngrid
\pagebreak
\section{supplemental Material}
Here we present fully quantum-mechanical calculations of the ground state of the system, and we show the quantum collapse to a state where the phases of light are correlated with the position of the BEC. Since the total Hilbert space is a tensor product of Hilbert spaces which comprise it, the numerical calculations are not easy to handle. Here, the position space consists of 64 points and the photonic cutoff is set to 10 for both modes of light; thus the dimension of the total Hilbert space is 7744.

We also calculate the scaling of the number of photons with the number of atoms in the superradiant phase.
\setcounter{equation}{0}
\setcounter{figure}{0}
\setcounter{table}{0}
\makeatletter
\renewcommand{\theequation}{S\arabic{equation}}
\renewcommand{\thefigure}{S\arabic{figure}}
\renewcommand{\bibnumfmt}[1]{[S#1]}
\renewcommand{\citenumfont}[1]{S#1}

\section{Ground state of the Hamiltonian}
\begin{figure}[htb!]
    \centering
\includegraphics[width=0.4\textwidth]{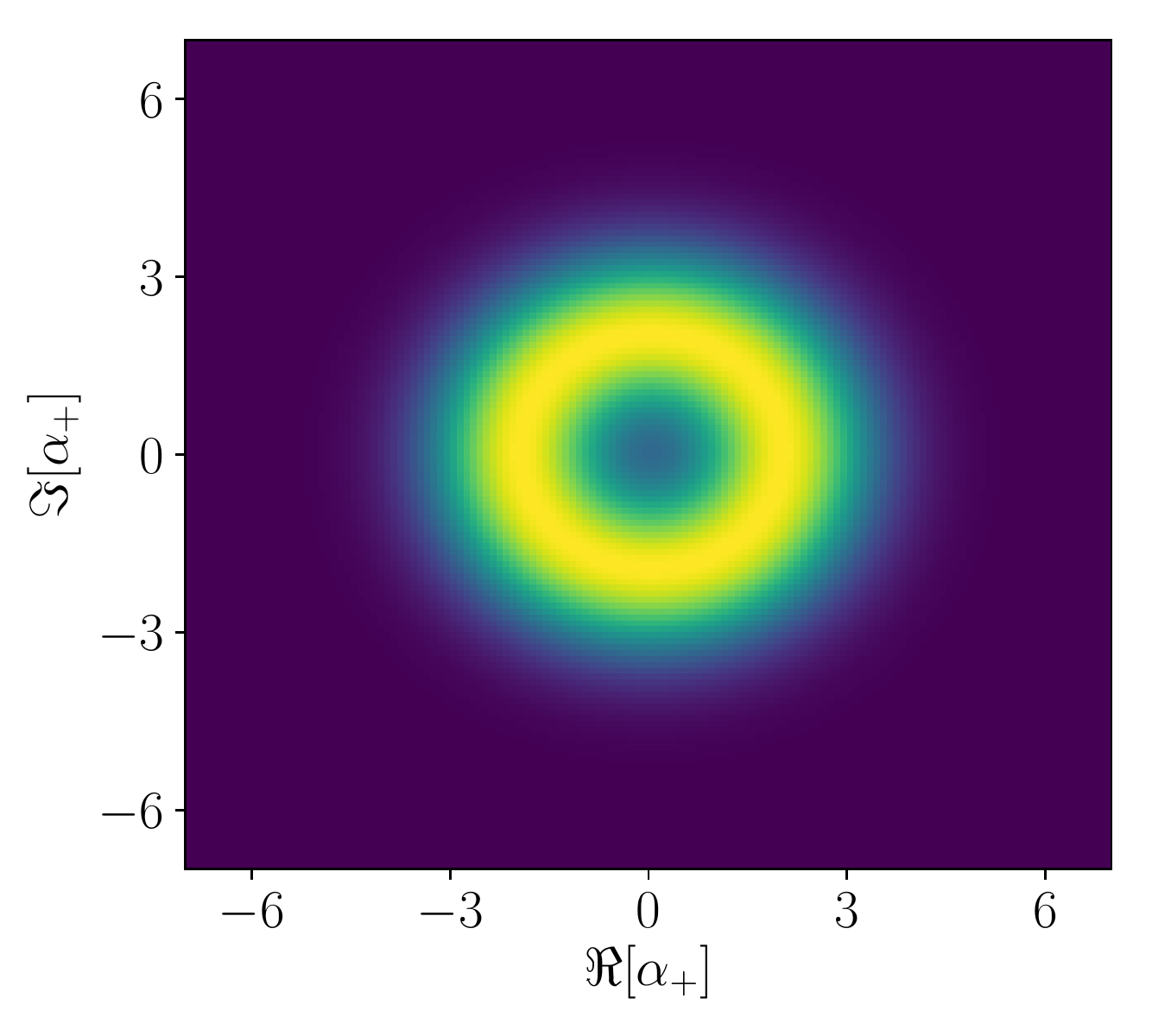}
\includegraphics[width=0.4\textwidth]{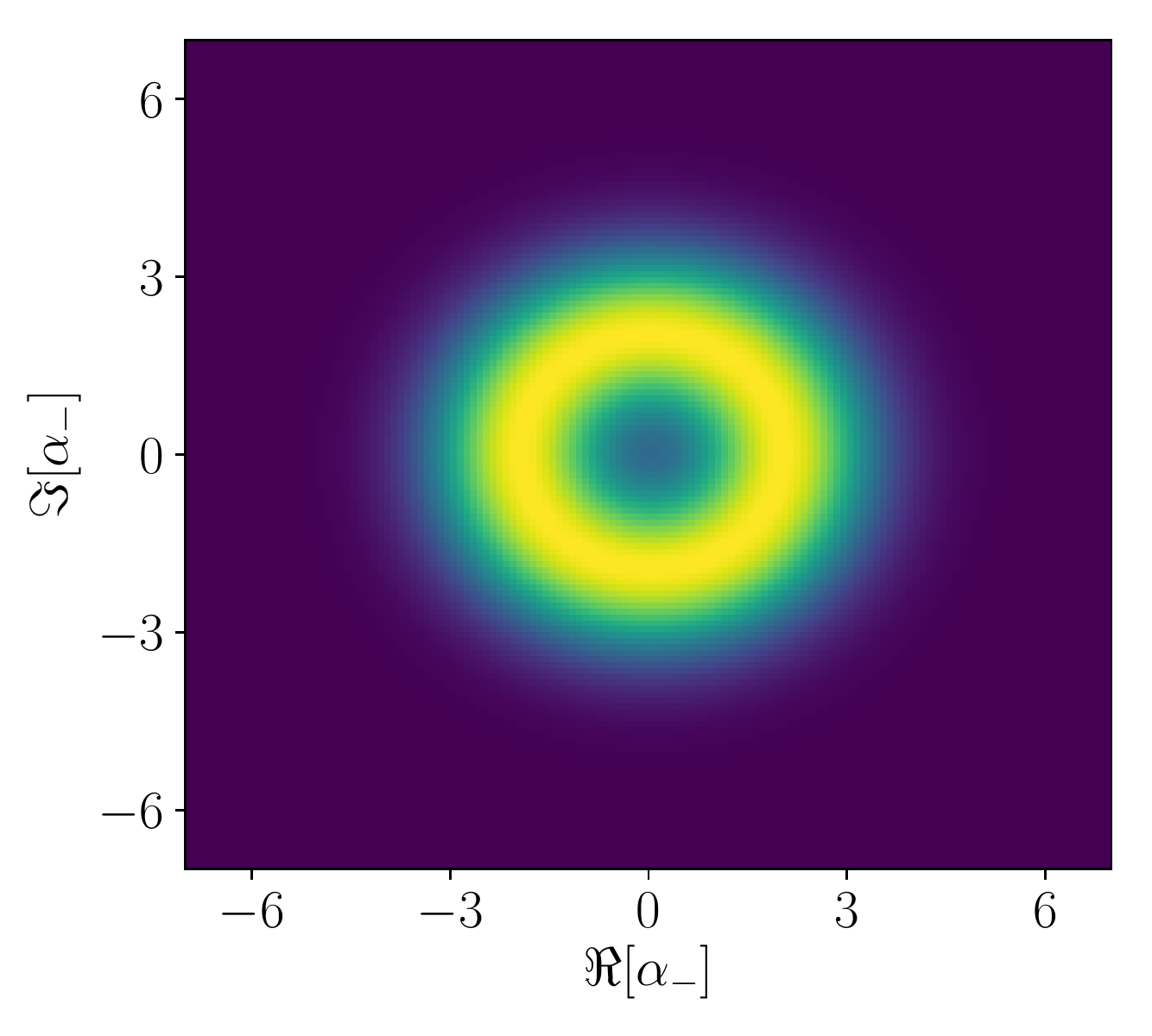}
\includegraphics[width=0.82\textwidth]{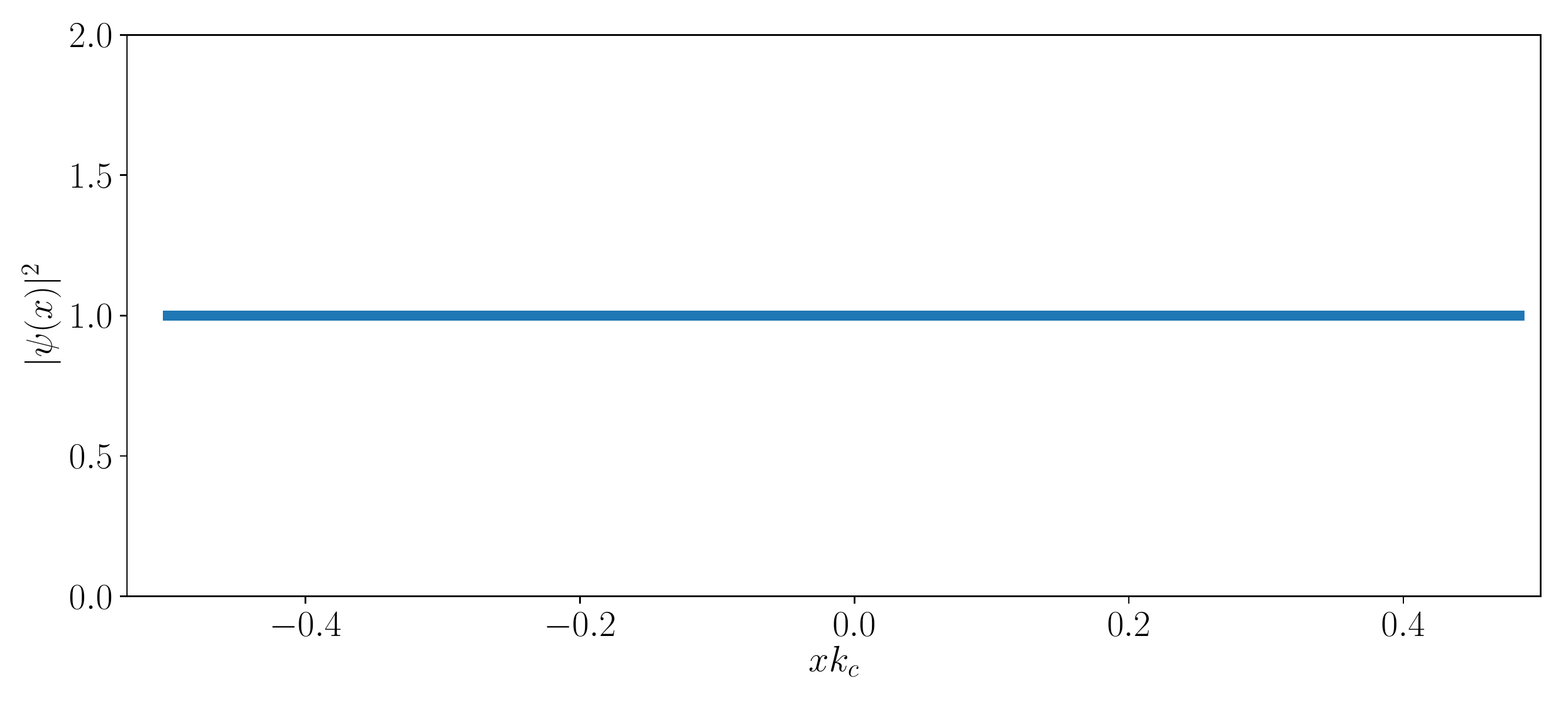}
\caption[ringstate]{The ground state of the Hamiltonian (\ref{eq:eff-H}) with $\theta=0$. The upper panel shows the $Q$ function of the light modes, while the lower panel shows the distribution of atoms along the cavity arm containing the BEC. Only one unit cell is showed.}
\label{fig:groundstate}
\end{figure}

The ground state of Hamiltonian (\ref{eq:eff-H}) is calculated by its exact diagonalization, and it can be written as
\begin{align}\label{eq:entint}
    | \Psi \rangle = \int \mathrm{d} x |x\rangle\otimes |\alpha e^{i x k_c }\rangle \otimes|\alpha e^{-i x k_c}\rangle,
    \end{align}
    where $\alpha$ is the amplitude of the electromagnetic field, and it depends on the number of atoms $N$ and the pump strength $\eta_0$. The ground state of the system with $\theta=0$ is presented in Fig.~\ref{fig:groundstate}.
\section{Quantum collapse}
\begin{figure}[htb!]
    \centering
\includegraphics[width=0.4\textwidth]{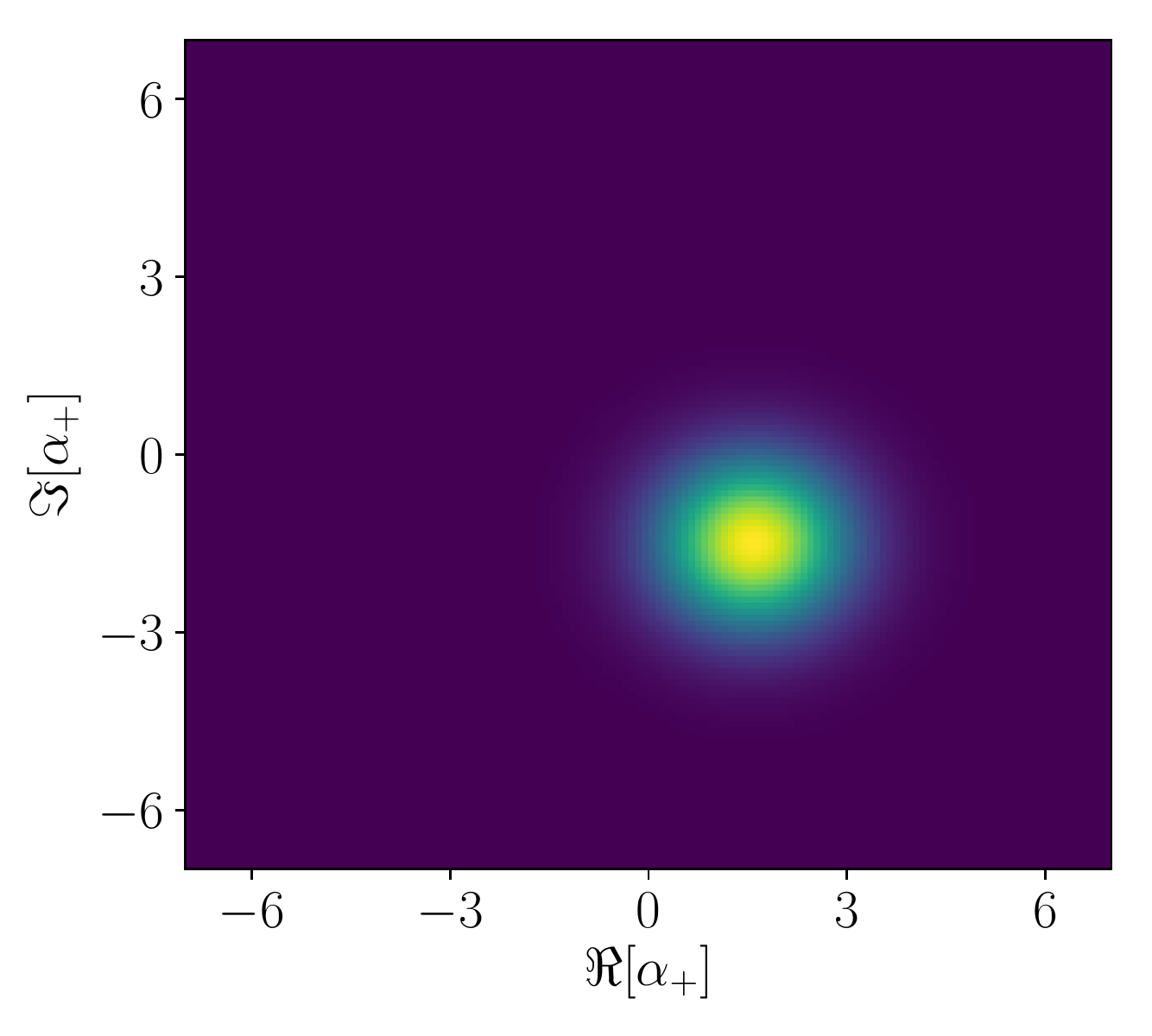}
\includegraphics[width=0.4\textwidth]{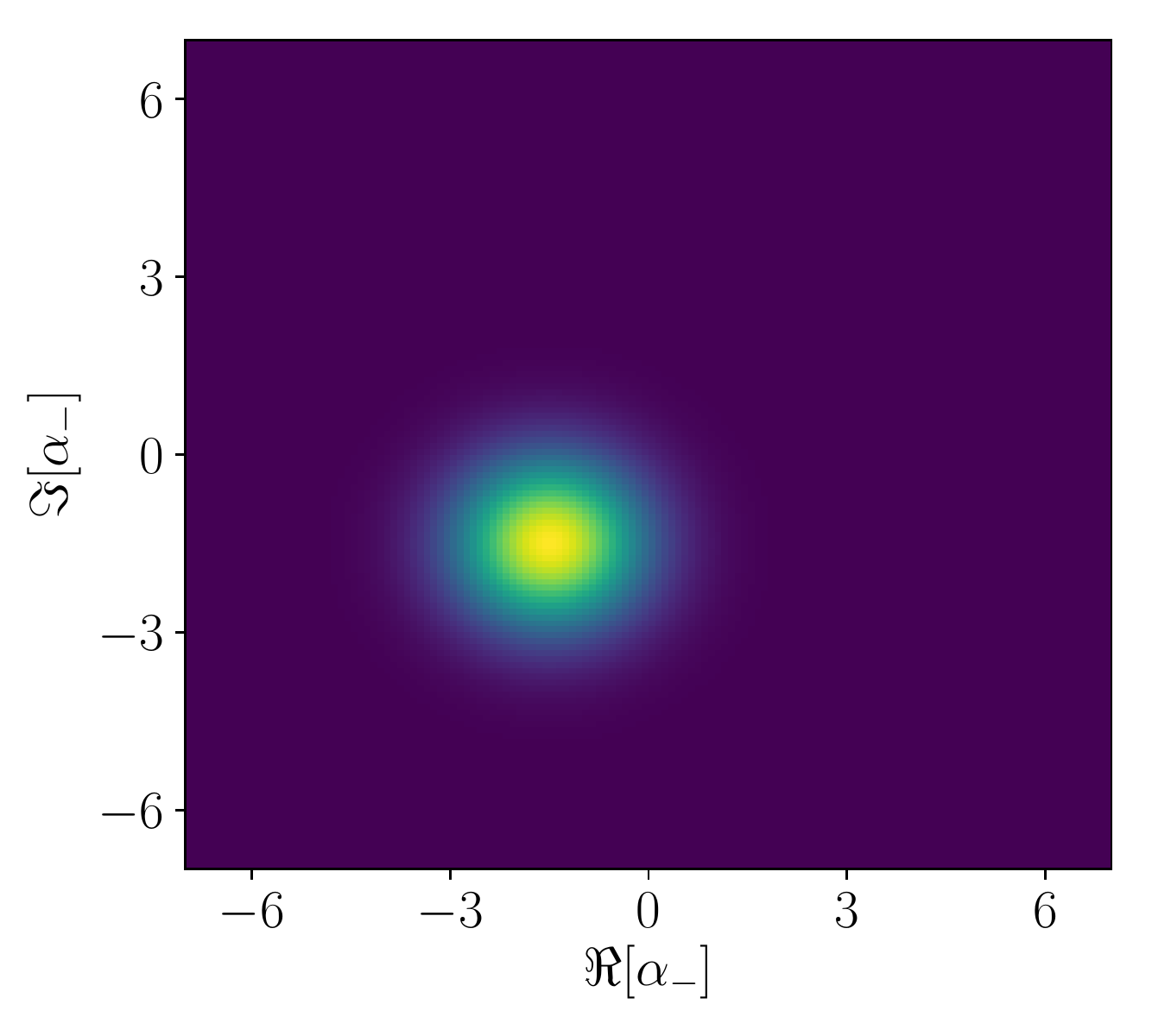}
\includegraphics[width=0.82\textwidth]{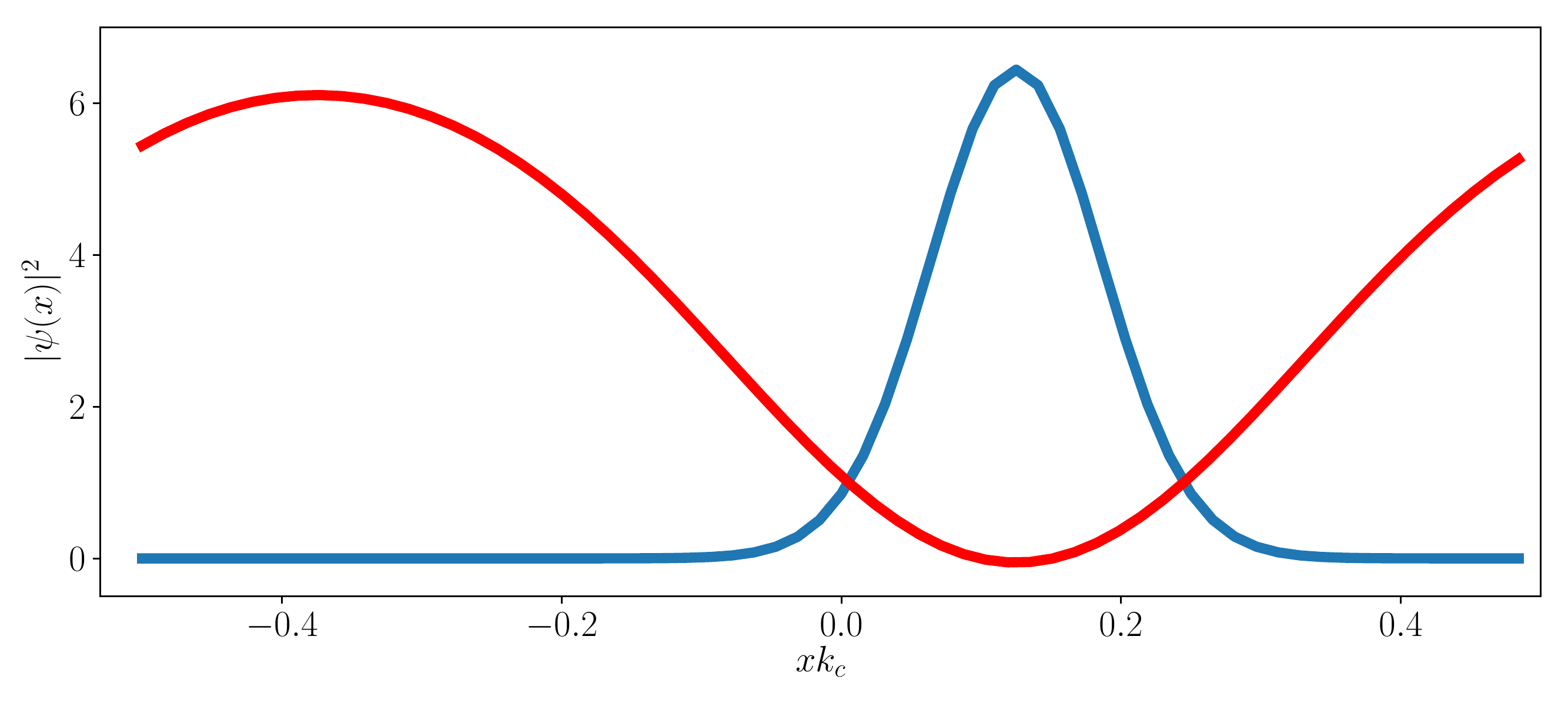}
\caption[ringstate]{The quantum collapse to a state with well defined phase correlated with the BEC position. The upper panel shows the $Q$ function of the light modes, while the lower panel shows the distribution of atoms along the cavity arm containing the BEC (blue curve) and the optical potential (red curve). Only one unit cell is showed.}
\label{fig:collapse}
\end{figure}

The quantum collapse (or the spontaneous symmetry breaking) of the state from Eq. (\ref{eq:entint}) is caused by decoherence induced, for instance, by losses of photons inside the ring cavity. Here we perform an artificial collapse by projecting the state from Eq. (\ref{eq:entint}) onto a coherent state. As a result, the state of the system is described by
\begin{align}
    | \psi \rangle = |x\rangle\otimes |\alpha e^{i x k_c}\rangle \otimes|\alpha e^{-i x k_c}\rangle.
\end{align}
The collapsed state is presented in Fig.~\ref{fig:collapse}.

\section{superradiance}
\begin{figure}[htb!]
    \centering
\includegraphics[width=0.9\textwidth]{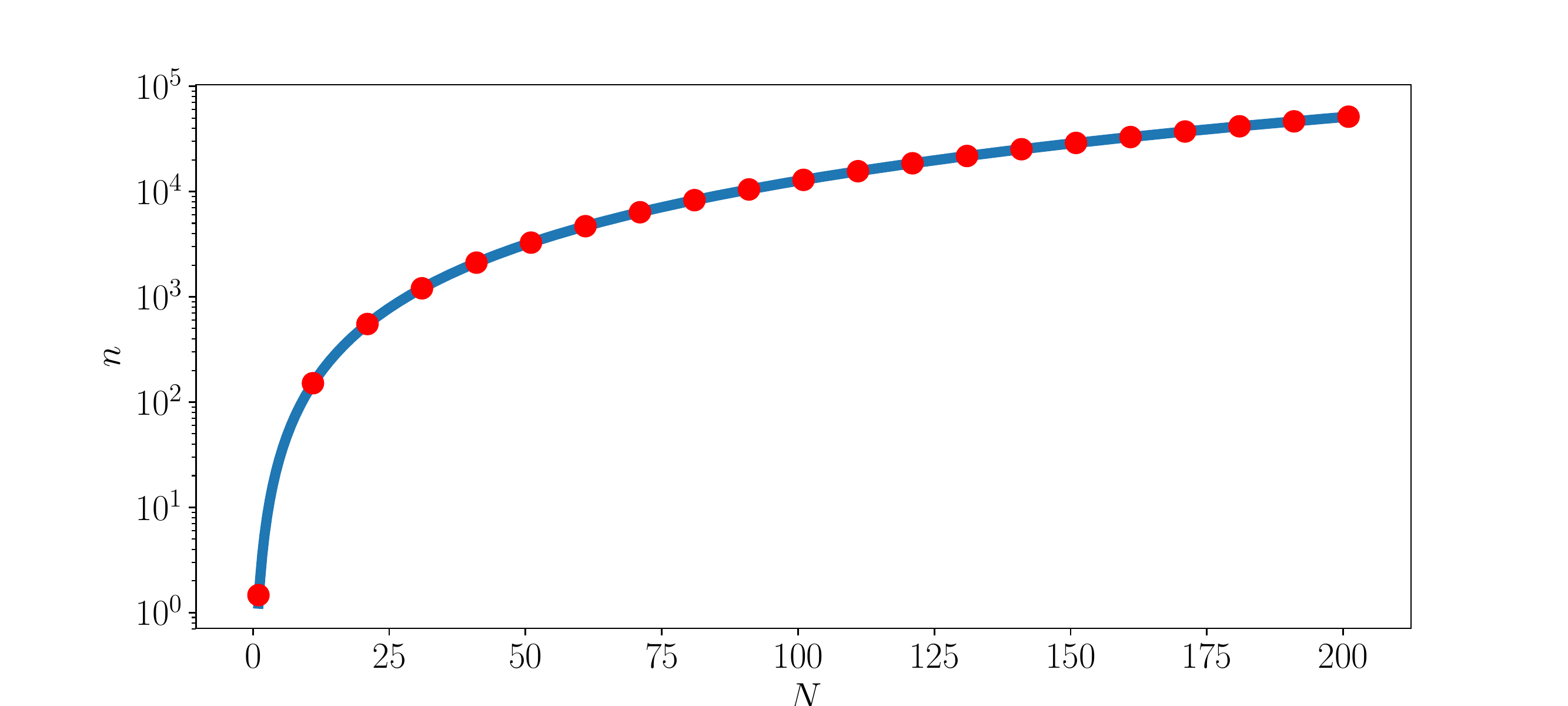}
\caption[scaling]{Superradiant intensity. Red points correspond to numerically calculated number of photons, and the solid blue line is the fit $n = a N^b$ with $a = 1.21778$ and $b =  2.00849$.}
\label{fig:scaling}
\end{figure}

The number of photons in the superradiant phase is determined by changing the number of atoms and self-consistently finding the steady state of the system. The scaling of the number of photons with a fitted dependence  $n = a N^b$ is presented in Fig.~\ref{fig:scaling}.

\end{document}